\documentclass[lettersize,journal]{IEEEtran}
\usepackage{cite}
\usepackage{amsmath,amsthm,amssymb,amsfonts}
\usepackage{mathrsfs}
\usepackage{algorithm}
\usepackage{algorithmicx}
\usepackage{algpseudocode}
\usepackage{listings}
\usepackage{graphicx}
\usepackage{textcomp}
\usepackage{multirow}
\usepackage{booktabs}
\usepackage{diagbox}
\usepackage{xcolor}
\usepackage{csquotes}
\usepackage{makecell}
\usepackage{array}
\usepackage[caption=false,font=normalsize,labelfont=sf,textfont=sf]{subfig}
\usepackage{stfloats}
\usepackage{url}
\usepackage{verbatim}
\begin{document}

\title{Stock Movement Prediction with Multimodal Stable Fusion via Gated Cross-Attention Mechanism}

\author{Chang Zong and Hang Zhou
\thanks{Chang Zong is with the School of Information and Electronic Engineering, Zhejiang University of Science and Technology, China (e-mail: zongchang@zust.edu.cn).}
\thanks{Hang zhou is with the Department of Finance Accounting and Economics, Business School of Nottingham University, China (e-mail: hang.zhou@nottingham.edu.cn)}
}

\markboth{Journal of \LaTeX\ Class Files,~Vol.~14, No.~8, August~2021}%
{Shell \MakeLowercase{\textit{et al.}}: A Sample Article Using IEEEtran.cls for IEEE Journals}

\maketitle

\begin{abstract}
The accurate prediction of stock movements is crucial for investment strategies. Stock prices are subject to the influence of various forms of information, including financial indicators, sentiment analysis, news documents, and relational structures. Predominant analytical approaches, however, tend to address only unimodal or bimodal sources, neglecting the complexity of multimodal data. Further complicating the landscape are the issues of data sparsity and semantic conflicts between these modalities, which are frequently overlooked by current models, leading to unstable performance and limiting practical applicability. To address these shortcomings, this study introduces a novel architecture, named Multimodal Stable Fusion with Gated Cross-Attention (MSGCA), designed to robustly integrate multimodal input for stock movement prediction. The MSGCA framework consists of three integral components: (1) a trimodal encoding module, responsible for processing indicator sequences, dynamic documents, and a relational graph, and standardizing their feature representations; (2) a cross-feature fusion module, where primary and consistent features guide the multimodal fusion of the three modalities via a pair of gated cross-attention networks; and (3) a prediction module, which refines the fused features through temporal and dimensional reduction to execute precise movement forecasting. Empirical evaluations demonstrate that the MSGCA framework exceeds current leading methods, achieving performance gains of 8.1\%, 6.1\%, 21.7\% and 31.6\% on four multimodal datasets, respectively, attributed to its enhanced multimodal fusion stability.\end{abstract}

\begin{IEEEkeywords}
Stock movement prediction, Multimodal fusion, Gated cross-attention
\end{IEEEkeywords}

\section{Introduction}
With the continual increase in the capitalization of the stock market, trading stocks has become an attractive investment instrument for many investors. The task of predicting stock movements, which focuses on forecasting future trends of a stock's prices, benefits the right selection of stocks and is of great significance for investment decisions \cite{feng2019temporal}. Stock movement prediction is a very difficult task. Traditional stock movement prediction methods apply machine learning techniques to perform data mining on financial indicators \cite{nelson2017stock} \cite{long2019deep}, social media documents \cite{nguyen2015sentiment}\cite{liu2019transformer}, or market relationships \cite{feng2019temporal} \cite{gao2021graph}. These methods employ a specific neural network encoder, which is suitable for a particular type of information, to represent stock features.

However, the price trend of a stock could be influenced by a variety of factors, and the extent of these factors' impacts can differ. These factors, collected from various data sources, are often presented in different modalities, including numerical indicators (e.g., prices, financial indicators), textual content (e.g., public opinions, news) and graph structures (e.g., industry relations, investment relations). How to effectively utilize the features of these multimodal factors to accurately predict stock price movements has always been a challenge. Recent multimodal approaches try to solve the stock movement prediction task considering two sources of information, such as the integration of the price and the text signal \cite{xu2018stock} \cite{ma2023multi}, or a blend of social media text and a correlation graph \cite{sawhney2020deep} \cite{li2021modeling}. These methods mostly apply multiple specific encoders subsequently for each modality and integrate features with simplified fusion strategies such as vector concatenation and conventional attention mechanism, without considering distinctions and conflicts between modalities. 

Meanwhile, there is still a distance between the task objectives and the actual demands. The latest multimodal deep learning methods focused only on predicting the rise and fall of stock prices, which is considered a binary classification problem \cite{xie2023wall} \cite{kaeley2023support} \cite{soun2022accurate}. However, slight fluctuation of stock prices is also an important phenomenon in real scenarios, usually called a sideways trend or consolidation, indicating that the forces of supply and demand are nearly equal before a new trend \footnote{https://www.investopedia.com/terms/s/sidewaystrend.asp}. However, some early studies involve multi-label classification with traditional deep learning models such as Long Short-Term Memory (LSTM) \cite{zhao2017time} and statistical methods such as random forests \cite{zhang2018novel}, there is still a research gap between multimodal fusion and multi-label stock prediction. Unifying these two settings has an unquestionably practical meaning for stock-trading scenarios.

Despite the more information introduced and the more complex architecture designed in previous works to approximate real-world situations while solving the stock movement prediction task, it still suffers from the inconsistency of the multimodal source, the deficient multimodal fusion strategies to make the accurate prediction, and the oversimplified task setting to meet real-world demands, which motivates our study to focus on multimodal feature fusion for fine-grained stock movement prediction. Specifically, the following challenges need to be addressed.

\textit{\textbf{Challenge 1}} involves the use of overly simplistic methods to tackle complex prediction tasks in current research. This includes the introduction of too few modalities and the application of basic fusion techniques, which fail to account for the diversity inherent in multimodal prediction. To predict stock movements, analysts typically use continuous financial indicators (such as opening and closing prices), textual information from various sources (including news, tweets, and blogs), and graph data reflecting market relationships (such as industry sectors and supply chains). However, existing research on stock movement prediction has been limited to unimodal \cite{nelson2017stock} \cite{long2019deep} \cite{liu2019transformer}  or bimodal \cite{kaeley2023support} \cite{zou2022astock} \cite{chen2021graph} analysis, resulting in predictions that often diverge significantly from actual market trends. Furthermore, multimodal fusion techniques significantly impact overall predictive accuracy. Current methods typically employ single-modal encoding followed by multimodal integration \cite{pawlowski2023effective}, often defaulting to simple vector concatenation for fusion \cite{fukui2016multimodal} \cite{wirojwatanakul2019multi} \cite{singh2019towards}, which fails to facilitate interactivity between different data types. Other studies employ co-attention mechanisms to fuse multimodal data \cite{zhang2018adaptive} \cite{yu2019deep}, yet often overlook the distinct contributions of each modality to the specific task. All these shortcomings of existing research drive us to conduct further studies on predicting stock trends with more data sources using a more optimized and efficient model.

\begin{figure}[htbp]
\centerline{\includegraphics[width=0.5\textwidth]{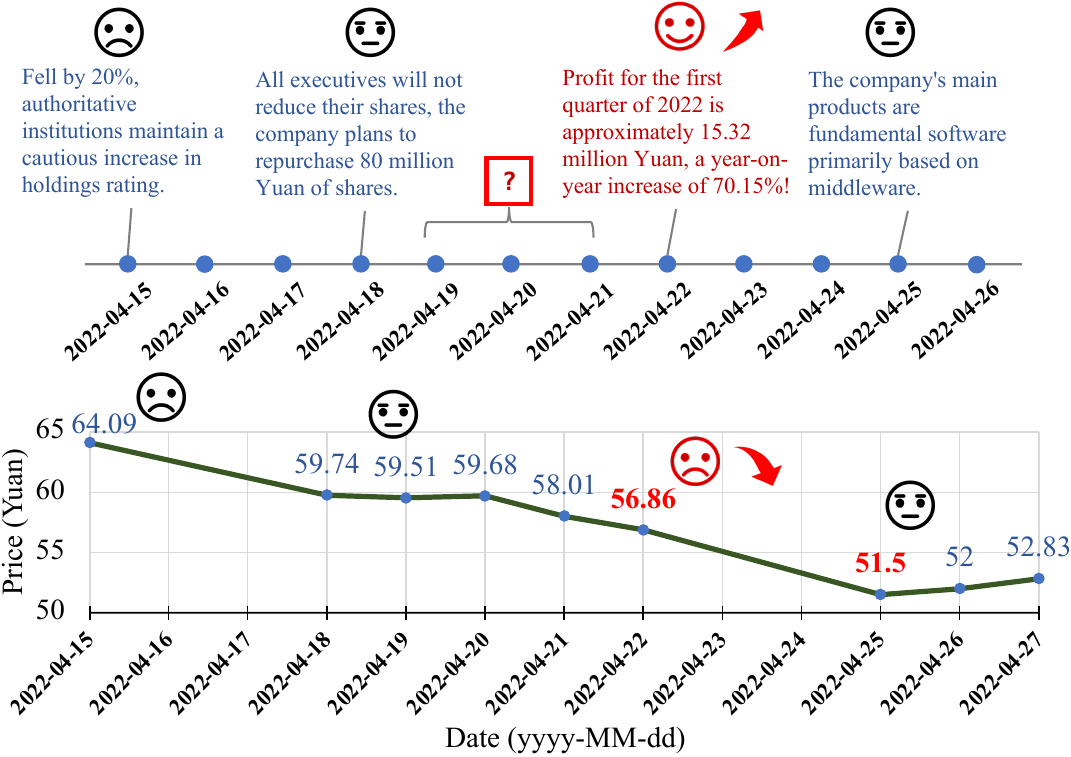}}
\caption{An empirical case from China's SSE STAR Market illustrates the challenges associated with the dataset of multimodal stock information. First, the dataset exhibits gaps in the news documents that align with price indicators, barring holiday periods. Additionally, a discernible semantic conflict is observed between the reported news content and concurrent price trajectories. These discrepancies are marked in red for emphasis.}
\label{example}
\end{figure}

\textit{\textbf{Challenge 2}} involves a data inconsistency problem while utilizing multiple data sources to address the prediction task. This includes data sparsity and semantic conflicts, which markedly impact the precision of predictions. As shown in Fig. \ref{example}, the data we gather from various sources frequently contains inconsistencies due to missing values, hindering temporal alignment across all modalities. Furthermore, semantic conflicts may arise when different modalities convey contradictory meanings. Traditional fusion techniques applied to these data sources can compromise or diminish predictive accuracy. Previous studies on multimodal stock movement prediction have simplified the issue by incorporating fewer sources of information \cite{xie2023wall} \cite{soun2022accurate} \cite{huynh2023efficient} \cite{daiya2021stock} or have neglected it altogether \cite{wang2023essential} \cite{he2021multi}. To address the problem of data sparsity among modalities, most previous methods concentrate on filtering and smoothing in the presence of missing values to handle uncertainty by a factorized inference model \cite{zhi2020factorized} or a heterogeneous graph model \cite{chen2020hgmf}. Another method \cite{ma2022multimodal} investigates that Transformer models are sensitive to incomplete modalities and to automatically search for an optimal strategy. Meanwhile, very few studies have focused on the problem of semantic conflicts between modalities. One way is to identify noise with a density estimation block based on the inherent correlation between modalities \cite{amrani2021noise}. Another work addresses the problem of the imbalance contribution of data sources by learning a network to determine the level of noise of each encoder \cite{mai2023multimodal}. Although these studies improve the performance of multimodal tasks by handling particular data problems between modalities, they cannot be universally applied to address this challenge and also increase the complexity of the overall architecture. This urges us to come up with a more general approach to solve these unstable fusion problems for stock movement prediction.

With the above challenges to be addressed and the weaknesses of the previous studies, we are prompted to explore the following question: \textbf{Is it possible to improve the performance of multimodal stock movement prediction in fine-grained task scenarios with a framework for stable fusion and efficient execution?}

To answer this question and address the above challenges, in this paper, we present \textbf{MSGCA}, a stock movement prediction framework with multimodal stable fusion via a gated cross-attention mechanism, which progressively integrates three types of modalities to predict stock trends on fine-grained labels (up, flat, and down). MSGCA considers three modalities including indicator sequences of stock prices, dynamic documents of sentiment and news, and a knowledge graph with industry relationship, each of which has either dynamic or static features. MSGCA performs the task of stock price movement prediction through three phases. To cope with \textit{\textbf{Challenge 1}}, we implement a trimodal encoder module as the first phase of MSGCA to handle these three heterogeneous modalities. We use a multilayer perceptron (MLP), a pre-trained large language model (LLM), and a graph attention neural network (GAT) to encode the information of indicators, documents, and a graph, respectively, obtaining the same dimension of latent representation across modalities. These unified features are fused via a gated cross-attention mechanism, which achieves feature interaction and perceives noisy information at the same time. To deal with \textit{\textbf{Challenge 2}}, a stable multimodal fusion module is implemented as our second phase. Our gated cross-attention network takes the indicator sequence (complete and accurate information) and the fused intermediate features as the primary modalities in two fusion stages to guide the fusion with other modalities, achieving stable integration without explicitly dealing with noise. MSGCA fuses the features of three modalities in succession using two gated cross-attention blocks to complete the overall multimodal fusion process. To further approach real-world demands, a fine-grained movement prediction module is implemented as the third module. We transform the fused features from the second module into probabilities corresponding to three movement trends (up, flat, down) for each stock using two MLP networks. These networks sequentially perform temporal and feature-based dimension reductions on latent embeddings to generate predictions. We constructed four multimodal stock trend prediction datasets based on existing benchmarks and the latest stock data that we collected. By comparing against leading unimodal and multimodal methods, our MSGCA model outperforms these baselines by margins of 8.1\%, 6.1\%, 21.7\%, and 31.6\% on four datasets, respectively.

Our contributions can be summarized as follows.
\begin{itemize}
    \item We implement, for the first time, a framework to solve multimodal fine-grained stock movement prediction, named MSGCA\footnote{The code of MSGCA and the data for training and evaluation are available at: https://github.com/changzong/MSGCA.}, which efficiently utilizes features from various modalities. This framework characterizes and integrates the indicator sequences of stock prices, the dynamic documents of sentiment and news, and the relational graph of industry sectors, to predict the up, flat, and down trends for stocks.
    \item We propose a method that can stably fuse multimodal features. This method implements a gated cross-attention mechanism, which allows a primary modality to guide the integration of other modalities step by step, effectively utilizing positive information and reducing noise from various modalities, different from all previous stock movement prediction methods.
    \item We present a trimodal encoder module as the initial phase of stock movement prediction that for the first time obtains latent representations of three heterogeneous modalities including indicator sequences (multiple types of price), dynamic documents (tweets and news), and a relational graph (industry sectors) simultaneously using MLP, LLM, and GAT models, respectively. 
    \item MSGCA is evaluated on four multimodal stock movement prediction datasets with real stock market information. The experimental results and analysis demonstrate that our method outperforms all baselines in terms of prediction accuracy and contributes to enhanced stability during multimodal fusion.
\end{itemize}

The remainder of the paper is organized as follows. First, we review the related work and make a brief comparison between our method and previous studies in Section \uppercase\expandafter{\romannumeral2}. Then we provide the preliminary and task formulation of our study in Section \uppercase\expandafter{\romannumeral3}. The implementation of our MSGCA stock movement prediction method is presented in \uppercase\expandafter{\romannumeral4}. Section \uppercase\expandafter{\romannumeral5} shows the experiments and analysis, followed by the conclusion and future work in Section \uppercase\expandafter{\romannumeral6}.

\section{Related Work}

\subsection{Stock Movement Prediction}
Stock movement prediction is challenging due to the market being highly stochastic and temporally dependent on predictions from chaotic data \cite{xu2018stock}. We summarize the existing stock movement prediction methods in Table \ref{existing}. Traditional methods focus on utilizing single-modal input with statistical \cite{nguyen2015sentiment} \cite{zhang2018novel} or simple deep learning methods \cite{feng2019temporal} \cite{nelson2017stock} \cite{long2019deep}. For example, an early work named JST \cite{nguyen2015sentiment} extracts topics and sentiments simultaneously using a linear kernel SVM and utilizes them for stock market prediction. MFNN \cite{long2019deep} learns an end-to-end model with integration of convolutional and recurrent neurons for feature extraction on financial time series samples. TGC \cite{feng2019temporal} proposes a temporal graph convolution, which jointly models the temporal evolution and the relation network of stocks. However, all of the above methods focus on a singular modality in one of indicators, text, or graph, thereby providing limited information for the multifaceted task. In addition, these works utilize LSTM or CNN variants that are limited in their representational capabilities. To address these problems, recent efforts on stock movement prediction focus on introducing more types of information \cite{xu2018stock} \cite{ma2023multi} \cite{he2021multi} and using modern deep learning methods such as the attention mechanism \cite{gao2021graph} \cite{soun2022accurate} \cite{huynh2023efficient} \cite{yoo2021accurate} \cite{feng2018enhancing} \cite{hsu2021fingat} and large language models \cite{xie2023wall} \cite{zou2022astock}. For example, TRAN \cite{gao2021graph} proposes a time-aware relational attention network for graph-based stock movement prediction to capture time-varying correlation strengths between stocks. MAC \cite{ma2023multi} proposes a multisource aggregated method that incorporates the numerical features, market-driven news, and the sentiments of their related stocks. ALSTM \cite{feng2018enhancing} complete the task using components that include feature mapping, LSTM, temporal attention, and prediction layer. SLOT \cite{soun2022accurate} learns the unified embeddings of prices and tweets from self-supervised learning and uses ALSTM for prediction. 

All the aforementioned methods focus on predicting the up and down trends of stock prices given various stock features. However, the flat trend of a stock is also important for prediction, which indicates that the forces of demand are nearly equal before a new period. Some previous studies aim to solve a fine-grained prediction but with single-modal input. In contrast, our proposed MSGCA tries to perform a stable multimodal fusion by introducing three types of modalities (indicator, documents, and graph) to solve the task with labels of up, flat and down, which fills the gap for fine-grained prediction with multimodal features as shown in Table \ref{existing}.

\begin{table}[htbp]
\caption{Summarization of Existing Stock Movement Prediction Methods.}
\centering
\begin{tabular}{l|c|c}
\bottomrule
\diagbox{Task Type}{Methods}{Input Type}  & Unimodal features & Multimodal features
\\ \hline      
Coarse-grained Prediction & \makecell[c]{JST \cite{nguyen2015sentiment} \\ CapTE \cite{liu2019transformer} \\ MFNN \cite{long2019deep} \\ ALSTM \cite{feng2018enhancing} \\ DTML \cite{yoo2021accurate}} &  \makecell[c]{TRAN \cite{gao2021graph} \\ StockNet \cite{xu2018stock} \\ MAC \cite{ma2023multi} \\ AStock \cite{zou2022astock} \\ ESTIMATE \cite{huynh2023efficient} \\ SLOT \cite{soun2022accurate}}
\\ \hline    
Fine-grained Prediction & \makecell[c]{Xuanwu \cite{zhang2018novel} \\ T-LSTM\cite{zhao2017time} \\ TASA\cite{sagala2020stock}}  & \textbf{MSGCA}
\\ \bottomrule
\end{tabular}
\label{existing}
\end{table}

\subsection{Multimodal Fusion Strategies}
Multimodal fusion is the process of combining collected data from various modalities for analysis tasks \cite{poria2017review}, and the fusion of multimodal data can provide surplus information with an increase in accuracy of the overall result \cite{d2015review}. Most early multimodal fusion strategies implement the framework of encoder and concatenation process, which focuses on using a vectorized concatenation operator to generate the final features \cite{fukui2016multimodal} \cite{wirojwatanakul2019multi} \cite{singh2019towards}. These simple fusion methods overlook the interference of noisy information while performing useful feature extraction \cite{fukui-etal-2016-multimodal}. Some other works utilize the attention and co-attention mechanism to simultaneously integrate information from all modalities \cite{zhang2018adaptive} \cite{lu2017knowing} \cite{yu2018beyond}, but ignore the differences between modalities for the targeted tasks. These methods involve on filtering noise while jointly learning the multimodal features. Modern approaches focus on more sufficient multimodal interactions \cite{yu2019deep} \cite{li2019beyond}. These studies employ self-attention and co-attention for intro-modal and inter-modal interaction, respectively, but suffer from using redundant information and architecture during the process, and further impair the efficiency of the framework \cite{zheng2023mmkgr}.

Considering the problem of noise interference, modality differences, and efficiency degeneration, our MSGCA focus on the stock movement prediction task, which achieves a stable multimodal fusion using a gated cross-attention mechanism guided by the main modality and features step-by-step during the fusion process.

\subsection{Cross- and Gated-Attention Mechanisms}
A powerful and robust model to represent features from multiple input sequences requires components that can effectively aggregate information relevant to the task from each source, and cross-attention is considered as an effective mechanism for feature integration \cite{rajan2022cross}. Existing cross-attention studies treat one sequence as query, and the other sequences as key and value to perform the encoder \cite{li2021selfdoc} \cite{jaegle2021perceiver} \cite{zhou2023transformer} \cite{tsai2022multimodal}. For example, IRENE \cite{zhou2023transformer} fuses textual and visual features with two bidirectional cross-attention blocks to achieve mutual integration between two types of input. Another work \cite{rajan2022cross} implements a tri-modal cross-attention architecture to perform cross-attention with combinatorial input from three sources. In addition, the gating mechanism has been shown to be essential for recurrent neural networks (RNNs) to achieve state-of-the-art performance \cite{jozefowicz2016exploring}. Gated linear unit (GLU) can reduce the problem of gradient vanishing by providing a linear path for gradients while maintaining non-linear capabilities \cite{dauphin2017language}. Gated attention unit to formulate attention and GLU as a unified layer, which allows the use of a much simpler attention mechanism and results in higher computational efficiency \cite{hua2022transformer}.

Following the above studies, our MSGCA employs cross-attention and gated-attention mechanisms to solve the feature integration problem from multiple sources to predict stock movement. Meanwhile, We distinguish stock-related information into primary and auxiliary features, and employ a fusion method involving three modalities through two consecutive gated cross-attention mechanisms to maintain high efficiency.

\section{Definition and formulation}
In this section, we provide some definitions and formulation of our task to better understand our method in this study.

\textbf{\textit{Definition 1: Multimodal source.}} The multimodal source $\mathcal{M}$ of a stock consists of various types of information as input to our stock movement prediction framework. In our study, we focus on three types of sources for each stock, represented as $\mathcal{M} = \{\mathcal{I}, \mathcal{D}, \mathcal{G}\}$, and each modality is formulated as follows.
\begin{itemize}
    \item \textbf{Indicator sequence}, expressed as $\mathcal{I} = \{i_1, i_2, ..., i_t\}$, where $i_t$ is the numerical indicator, such as the close price of a stock produced at timestamp $t$.
    \item \textbf{Dynamic document}, represented as $\mathcal{D} = \{d_1, d_2, ..., d_t\}$ and $d_t = \{p^1_t, p^2_t, ..., p^n_t\}$, where $d_t$ is the set of tweets or news in the published time stamp $t$, and $p^n_t$ is the $n$-th paragraph that composes $d_t$.
    \item \textbf{Relational graph}, shows the intra-industry relationships between companies \cite{hsu2021fingat} \cite{ye2021multi}, denoted as $\mathcal{G} = \{\mathcal{E}, \mathcal{R}, \mathcal{U}\}$, where $\mathcal{E}$ is the set of entities and $\mathcal{R}$ is the set of relations. $\mathcal{U} = \{(e_s, r, e_d)|e_s,e_d \in \mathcal{E}, r \in \mathcal{R} \}$ is the set of triplets in the graph $\mathcal{G}$, where $e_s$, $e_d$ and $r$ are the start entity, the end entity, and the relation, respectively.
\end{itemize}
According to the data we collect from real-world sources, the indicator sequence and dynamic documents of a stock contain a time dimension. Whereas, the relational graph is static.

\textbf{\textit{Definition 2: Multimodal fused feature.}} The multimodal fused feature $\textbf{\textit{x}}$ of a stock is transformed from its multimodal sources $\mathcal{M}$ by an encoder-fusion process, denoted $\textbf{\textit{x}} = \textbf{\textit{v}}_i \circ \textbf{\textit{v}}_d \circ \textbf{\textit{v}}_g$, where $\circ$ is the multimodal fusion method. The indicator feature $\textbf{\textit{v}}_i$, the document feature $\textbf{\textit{v}}_d$, and the graph feature $\textbf{\textit{v}}_g$ are learned from their corresponding encoders.

\textbf{\textit{Definition 3: Fine-grained movement prediction.}} Fine-grained movement prediction is to generate the probability for each trend label $l$ from the set of labels $L$ using the fused multimodal feature of a stock $\textbf{\textit{x}}$, denoted $Pr(y \in L | \textbf{\textit{x}}) = f(\textbf{\textit{x}})$, where $f$ is the prediction function, and $L = \{up, flat, down\}$.

\textbf{\textit{Objective: }}This study aims to predict the trend of price movement of a stock in the timestamp $t+1$, given the multimodal source of this stock from timestamp $1$ to timestamp $t$. The problem formulation is given as:
\begin{itemize}
    \item The input is the multimodal source of a stock $\mathcal{M} = \{\mathcal{I}, \mathcal{D}, \mathcal{G}\}$ which contains an indicator sequence $\mathcal{I}$, a dynamic document $\mathcal{D}$ and a static graph $\mathcal{G}$.
    \item The output is a probability distribution $p=Pr(y \in L | \textbf{\textit{x}})$ for the set of target trend labels $L$ generated by an encoder-fusion-decoder process.
\end{itemize}

\section{Methodology}

\subsection{Framework of MSGCA}
\begin{figure*}[!ht]
\begin{center}
    \includegraphics[width=1\textwidth] {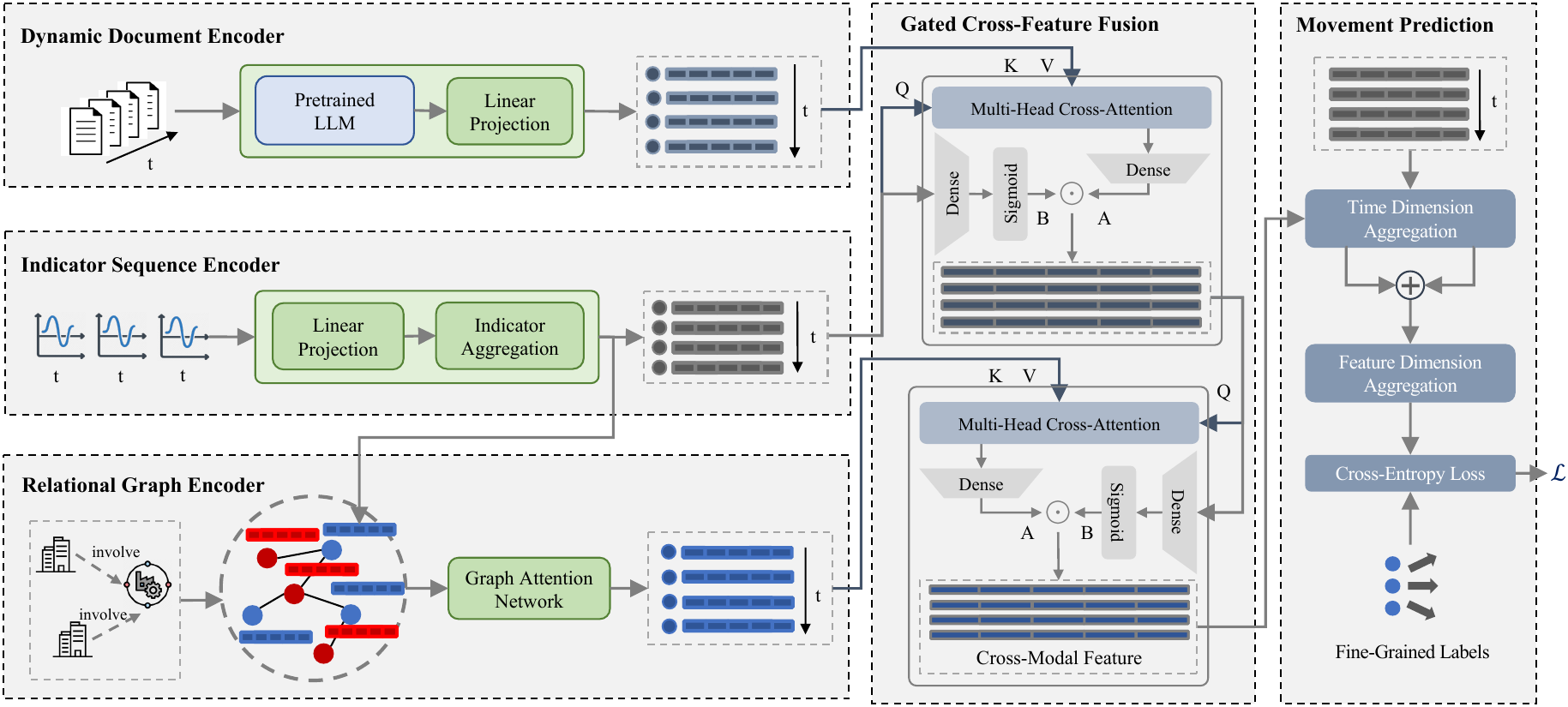}
\caption{Framework of MSGCA. Indicator sequences, dynamic documents, and a relational graph are encoded with a linear projection, a pre-trained large language model (LLM), and a graph attention network (GAT), respectively. The encoded features are stably fused with two gated cross-attention networks subsequently, guided by the primary modality (price indicators) and the intermediate features. The movement predictor transforms the fused features into the probabilities for each trend labels by performing aggregations along the temporal and feature dimensions with two MLP networks. All parameters in this framework are optimized by backpropagation with the overall loss $\mathcal{L}$.}
\label{framework}
\end{center}
\end{figure*}

The framework of our proposed method \textbf{MSGCA} is shown in Fig. \ref{framework}. The procedure is structured into three phases, adhering to an encoder-fusion-decoder architecture. During the multimodal encoding phase, projections from three modalities are transposed into a cohesive latent space through employing various representational methods as encoders. The gated cross-feature fusion phase implements a stable multimodal integration strategy to synthesize a unified feature representation through a pair of gated cross-attention networks. The movement prediction phase compresses both temporal and feature dimensions, aligning them with the granularity of the trend labels, thereby enabling precise prediction of stock movement at timestamp $T+1$. In Table \ref{notation}, we summarize the notations used in our framework.

\begin{table}[htbp]
\caption{Notations Used in Our Method.}
\centering
\begin{tabular}{l|l}
\bottomrule
\textbf{Notation} & \textbf{Meaning}
\\ \hline     
$\textbf{s}_c$, $\textbf{s}_o$, $\textbf{s}_h$ & indicator sequences of closed, open, and highest price  \\
$\textbf{s}_d$ & input sequence of dynamic documents
\\
$\textbf{v}_{i}$, $\textbf{v}_{d}$, $\textbf{v}_{g}$ & latent representations of three modalities
\\
$\mathcal{F}(\cdot)$, $\mathcal{T}(\cdot)$ & functions of filling zero vectors and getting text embeddings
\\
$\textbf{v}_{d}^{'}$ & document embedding generated from an LLM
\\
$\textbf{W}$, $b$, $b^{'}$ & model weights and bias to be learned
\\ 
$\textbf{H}$, $\textbf{h}$ & updated stock representation after feature fusion
\\ 
$\textbf{Q}$, $\textbf{K}$, $\textbf{V}$ & query, key, and value for cross-attention layers
\\ 
$K$, $M$ & number of heads for multi-head GAT and cross-attention 
\\
$d$, $d^{'}$ & embedding dimensions of learnable parameters and features
\\
$\textbf{p}$, $\textbf{l}$ & multi-class probability output and label vectors
\\
$\mathcal{L}$ & overall loss of the framework
\\ \bottomrule
\end{tabular}
\label{notation}
\end{table}

\subsection{Multimodal Source Encoding}
Analysis of stock movement is based on the idea that historical market behavior can inform traders about potential future trends. Conventional methods for stock movement prediction focus on employing statistical models on a range of financial indicators, including open prices, highest prices, lowest prices, closed prices, and trading volumes \cite{nelson2017stock} \cite{nguyen2015sentiment} \cite{zhang2018novel}. Recent deep learning approaches are apt to introduce more types of data sources, such as tweet text, and market relational data \cite{gao2021graph} \cite{soun2022accurate} \cite{hsu2021fingat}. The above methods encompass the use of indicators, documents, and graphs, which inspires us to concentrate our efforts on the management of these three distinct data sources. We implement a trimodal encoding module as the first phase of MSGCA to concurrently encode each type of multimodal information into a unified latent dimensional space. Each encoder is described as follows.

\textit{1) Indicator sequence encoding: }In our framework, the indicator sequence of a stock is considered its primary modality, which is characterized by data completeness and continuity. MSGCA applies three fundamental indicators, including closed prices, open prices, and highest prices, to demonstrate its proficiency in processing sequences of multiple indicators. Each indicator category is mapped into a $d$-dimensional latent feature space using a specified linear layer. Then, all three indicator features are concatenated and transformed into a unified feature sequence to facilitate the upcoming multimodal fusion. The encoding process for indicator sequences is formulated as the following equations:

\begin{gather}
\textbf{v}_{i} = (\textbf{v}_{c} \oplus \textbf{v}_{o} \oplus \textbf{v}_{h})\textbf{W}_i+b_i \\
\textbf{v}_{c} = \textbf{s}_{c}\textbf{W}_{ic}+b_{ic} \\
\textbf{v}_{o} = \textbf{s}_{o}\textbf{W}_{io}+b_{io} \\
\textbf{v}_{h} = \textbf{s}_{h}\textbf{W}_{ih}+b_{ih}
\label{eq:eq1}
\end{gather} where $\textbf{s}_{c}$, $\textbf{s}_{o}$, and $\textbf{s}_{h} \in \mathbb{R}^{t \times 1}$ are indicator sequences of closing, open, and highest prices, respectively, with time dimension $t$. $\textbf{v}_{c}$, $\textbf{v}_{o}$, and $\textbf{v}_{h} \in \mathbb{R}^{t \times d}$ are the latent embeddings of theirs corresponding indicators, $\oplus$ is the concatenation operator, $\textbf{W}_i \in \mathbb{R}^{3d \times d}$ is the learnable weight of the linear projection, and $\textbf{v}_{i} \in \mathbb{R}^{t \times d}$ is the overall feature of an indicator sequence.

\textit{2) Dynamic document encoding: } Social media content and news reports on a stock can reveal investor sentiment and corporate developments, which are factors to influence future stock valuations. These sources of textual information are presented as dynamic documents that require feature extraction for the purposes of our predictive analysis. However, due to the scarcity of data acquisition channels, documents pertaining to a particular stock do not invariably accompany its daily market activity. Moreover, Large Language Models (LLMs) such as ChatGPT \footnote{https://chat.openai.com/} and LLaMA \cite{touvron2023llama}, have excelled in various natural language tasks, which inspires us to employ a pre-trained LLM to perform the document encoding step. In our framework, we use text-embedding-ada-002 \footnote{https://platform.openai.com/docs/guides/embeddings} from OpenAI as our parameter-fixed LLM to convert textual documents into high-dimensional latent embeddings. Those vectorized representations are then subjected to a dimensionality alignment with the indicator modality through a linear layer. Furthermore, to address instances where document timestamps are missing, we populate the temporal sequence with zero vectors to maintain consistency with the corresponding indicator features. The encoding process for dynamic documents is described as the following equations:
\begin{gather}
\textbf{v}_d = \mathcal{F}(\textbf{v}_{d}^{'}\textbf{W}_d + b_d) \\
\textbf{v}_{d}^{'} = \mathcal{T}(\textbf{s}_d, LLM)
\label{eq:eq2}
\end{gather} where $\mathcal{T}(\cdot)$ is the function to get the latent text embeddings $\textbf{v}_d^{'} \in \mathbb{R}^{t^{'} \times 1536}$ from the given document sequence $\textbf{s}_d$ by calling the API service provided by OpenAI, using a pre-trained large language model $LLM$, and $\textbf{W}_d \in \mathbb{R}^{1536 \times d}$ is the learnable weight of the linear projection. The dynamic document feature $\textbf{v}_d \in \mathbb{R}^{t \times d}$ is aligned to its respective indicator sequence through a zero vector filling function $\mathcal{F}(\cdot)$.

\textit{3) Relational graph encoding: } Relationships between stocks often reflect sectoral and financial influences, which are useful in predicting stock price movement. We collect data describing the interconnections between stocks and their respective sectors and construct a relational graph to serve as our graph modality. As industry-specific fluctuations can concurrently affect stocks within the same sector, we employ a graph attention network (GAT) \cite{velickovic2017graph} as the graph encoder to capture these correlated variations. Meanwhile, to make the information related to each stock propagate through the graph, nodes representing individual stocks are initialized with vectors corresponding to indicator features at discrete time intervals. This operation facilitates the extension of the temporal dimension within the graph modality, ensuring alignment with the other two modalities. The relational graph encoding process is denoted by the following equations:
\begin{gather}
\textbf{v}_g = [\textbf{v}_g^1, \textbf{v}_g^2, ..., \textbf{v}_g^t] \\
\textbf{v}_g^t = \sigma \left(\frac{1}{K} \sum_{k=1}^{K} \sum_{j \in \mathcal{N}} \alpha_j^k \textbf{h}_j^{t} \textbf{W}^k\right) \\
\textbf{h}_j^{t(0)} = \textbf{v}_{i}^{j,t}
\label{eq:eq3}
\end{gather} where the overall graph embedding $\textbf{v}_g \in \mathbb{R}^{t \times d}$ is composed of embeddings from each timestamp $\textbf{v}_g^t \in \mathbb{R}^{1 \times d}$. $K$ is the number of heads to perform multi-head attention. $\mathcal{N}$ is the neighborhood of the central node. $\alpha_j^k$ is the attention coefficient for each neighboring node in each head. $\textbf{W}^k \in \mathbb{R}^{d \times d}$ is the learnable weight matrix, and $\textbf{h}_j^{t} \in \mathbb{R}^{1 \times d}$ is the hidden embedding of node $j$ in timestamp $t$ from the previous layer of GAT. $\sigma$ is an activity function. The initial embedding $\textbf{h}_j^{t(0)}$ of the node $j$ in timestamp $t$ is its indicator feature $\textbf{v}_{i}^{j,t} \in \mathbb{R}^{1 \times d}$.

\subsection{Gated Cross-Feature Fusion}
Data sparsity and semantic conflicts within multimodal fusion significantly compromise stock movement prediction efficacy, as described in our secondary challenge. Meanwhile, stock indicator sequences typically exhibit superior data integrity compared to other data modalities and ought to be prioritized as the primary input in our analysis. In the second module of MSGCA, we introduce an innovative gated cross-attention network designed to integrate three distinct modalities. This fusion is orchestrated by leveraging primary and stable features to guide the fusion process, distinguishing our approach from all other existing stock movement prediction methods.

The gated cross-attention network is inspired by studies on efficient transformers \cite{hua2022transformer}, cross-attention fusion \cite{li2021selfdoc} and gated networks \cite{dauphin2017language}. To integrate features from two modalities, the network initiates by executing a deeply interactive feature fusion via multi-head cross-attention mechanisms, thereby acquiring features in an unstable state. Then, using a stable feature as a discriminator, the network executes a gated layer operation to select the useful information from the unstable features for the subsequent processing stages. The details of this procedure are illustrated as follows.

\textit{1) Unstable cross-attention fusion: }During the fusion of indicator sequences with dynamic documents, a multi-head cross-attention network is employed to realize the deep integration of the two modalities. We configure the indicator features as queries, and the document features as keys and values, thereby executing a multi-head cross-attention process. This fusion process is described in the following equations:
\begin{gather}
\textbf{H}_{i,d}^l = softmax\left(\frac{1}{M} \sum_{m=1}^{M} \frac{\textbf{Q}_m^l (\textbf{K}_m^l)^\top}{\sqrt{d^{'}}}\right) \textbf{V}_m^l \\
\textbf{Q}_m^l = \textbf{H}_i^l\textbf{W}_m^Q,
\textbf{K}_m^l = \textbf{H}_d^l \textbf{W}_m^K, \textbf{V}_m^l = \textbf{H}_d^l \textbf{W}_m^V
\label{eq:eq4}
\end{gather} where $\textbf{H}_{i,d}^l$ is the unstable features from the fusion of indicator sequences and dynamic documents using multi-head cross-attention. $M$ is the number of heads. $d^{'} = M \times d$. $\textbf{W}_m^Q$, $\textbf{W}_m^K$, and $\textbf{W}_m^V \in \mathbb{R}^{d \times d'}$ are learnable weights of the query $\textbf{H}_i^l$, key $\textbf{H}_d^l$, and value $\textbf{H}_d^l$ from the $l$-th layer. 

\textit{2) Stable gated feature selection: }The previously integrated features exhibit instability as a result of data sparsity and semantic conflicts, as described in Fig. \ref{example}. So a step of latent feature selection is operated using a gated network architecture, which is guided by indicator sequences (our primary modality), in order to extract stable multimodal features. The gated feature selection is formulated as the following equations:
\begin{gather}
\textbf{H}_{i,d} = \textbf{H}_a \odot \textbf{H}_b \\
\textbf{H}_a = \textbf{H}_{i,d}^l \textbf{W}_a + b\\
\textbf{H}_b = Sigmoid(\textbf{H}_i \textbf{W}_b + b^{'})
\label{eq:eq5}
\end{gather} where $\textbf{H}_a$ and $\textbf{H}_b$ are the unstable-fused feature and main modality feature, respectively. $\textbf{W}_a \in \mathbb{R}^{t \times d}$ and $\textbf{W}_b \in \mathbb{R}^{t \times d}$ are learnable weights. $b \in \mathbb{R}^{d}$ and $b^{'} \in \mathbb{R}^{d}$ are learnable biases. $\odot$ is the element-wise product operator between matrices. Finally, we get the stable fused feature $\textbf{H}_{i,d}$ from indicator sequences and dynamic documents.

\textit{3) Expansion to the third modality: }Similarly, the fusion of the third modality harnesses the stable characteristic derived from the previous gated cross-attention block to guide the subsequent stage of fusion. To integrate information from the graph modality, a secondary cross-attention network is introduced as the following equations:
\begin{gather}
\textbf{H}_{i,d,g}^l = softmax\left(\frac{1}{M} \sum_{m=1}^{M} \frac{\textbf{Q}_m^l (\textbf{K}_m^l)^\top}{\sqrt{d^{'}}}\right) \textbf{V}_m^l, \\
\textbf{Q}_m^l = \textbf{H}_{i,d}^l\textbf{W}_m^Q, \textbf{K}_m^l = \textbf{H}_g^l \textbf{W}_m^K, \textbf{V}_m^l = \textbf{H}_g^l \textbf{W}_m^V
\label{eq:eq6}
\end{gather} where $\textbf{H}_g^l$ is the hidden features of the graph modality. $\textbf{H}_{i,d,g}^l$ is the unstable features which fuse three modalities. Finally, the stable fused feature of a stock derived from three modalities can be computed as the following equations:
\begin{gather}
\textbf{H}_{i,d,g} = \textbf{H}_a \odot \textbf{H}_b, \\
\textbf{H}_a = \textbf{H}_{i,d,g}^l \textbf{W}_a + b,\\
\textbf{H}_b = Sigmoid(\textbf{H}_{i,d} \textbf{W}_b + c)
\label{eq:eq7}
\end{gather} where $\textbf{H}_{i,d,g}$ is the final stable feature across the information of three modalities guided by the feature $\textbf{H}_{i,d}$ from the previous fusion stage. It is worthy to acknowledge that the scalability of our gated cross-attention mechanism extends to the integration of additional modalities within a sequential fusion paradigm.

\subsection{Fine-Grained Movement Prediction}
To predict stock movements, we should convert the integrated features into trend labels. In the third module of MSGCA, we compress the temporal and feature dimensions of stocks with two multi-layer perceptron (MLP) networks to map the fused features into the label space. The detailed steps are described below.

\textit{1) Time dimension aggregation: } Unlike other sequential methods such as LSTM \cite{sak2014long} and self-attention \cite{vaswani2017attention} applied to reduce the time dimension when calculating the overall representations of sequences, we directly use an MLP network for efficient dimension reduction. Meanwhile, considering that we only employ stable features as the gate to choose useful information when performing gated cross-attention, we further concatenate the indicator encoding features and the cross-modality features in this stage, to provide a chance to make the prediction directly with the primary features and avoid the potential usage of extreme noisy information from other modalities. This process is formulated as the following equations:
\begin{gather}
\textbf{h} = \textbf{h}_{i,d,g} \oplus \textbf{h}_i \\
\textbf{h}_{i,d,g} = MLP_t(\textbf{H}_{i,d,g})\\
\textbf{h}_i = MLP_t(\textbf{H}_i)
\label{eq:eq8}
\end{gather} where $MLP_t(\cdot)$ is the time aggregation network that gradually reduces the time dimension from $t$ to $1$ using three linear layers followed by three activation functions. $\textbf{H}_{i,d,g}$ and $\textbf{H}_i$ are cross-modal features and indicator sequence features, respectively. $\oplus$ is the concatenation operator to obtain the features with the time dimension removed $\textbf{h} \in \mathbb{R}^{2 \times d}$.

\textit{2) Feature dimension aggregation: } Our last step is to project the overall sequence features in the latent space onto each trend label to complete the task. We implement another MLP network to gradually reduce the feature dimension from $2 \times d$ to $3$, which is the number of trend categories, to represent the probabilities of each label. This step is illustrated as the following equation:
\begin{gather}
\textbf{p} = MLP_f(\textbf{h})
\label{eq:eq9}
\end{gather}, where $MLP_f(\cdot)$ is the feature aggregation network to gradually reduce feature dimension from $d$ to $3$ using the same structure as $MLP_t(\cdot)$. $\textbf{p} \in \mathbb{R}^{3}$ is the probability vector with feature dimension reduced from  $\textbf{h}$.

\textit{3) Multi-label loss function: } With the probability for each label (up, flat, and down) is obtained, a multi-class cross-entropy function is employed to calculate the loss of the overall process, which is formulated as the following equation:
\begin{gather}
\mathcal{L} = -\frac{1}{N} \sum_{n=1}^N \sum_{c=1}^{C}log\frac{exp(\textbf{p}_{n,c})}{\sum_{i=1}^C exp(\textbf{p}_{n,i})} \textbf{l}_{n,c}
\label{eq:eq10}
\end{gather} where $N$ is the number of batches. $\textbf{p}_{n,c}$ is the probability of the class label $c \in \{1, 0, -1\}$ for the $n$-th batch, where label $1$, $0$, and $-1$ stands for the trends of up, flat, and down, respectively. $\textbf{l}_{n,c}$ is the label of the $n$-th sample and the $c$-th class from the label vector $\textbf{l}$. $\mathcal{L}$ is the mean value of the overall loss.

\subsection{Pseudocode of Training MSGCA}
To conclude the training process of MSGCA with the above phases, the pseudocode is illustrated as Algorithm \ref{algorithm}.
\begin{algorithm}
    \caption{Pseudocode of Training MSGCA}
    \label{algorithm}
    \textbf{Input:} Indicator sequence: $\mathcal{I}=\{i_1, i_2, ..., i_t\}$, Dynamic documents: $\mathcal{D} = \{d_1, d_2, ..., d_t\}$, Relational graph: $\mathcal{G} = \{\mathcal{E}, \mathcal{R}, \mathcal{U}\}$ \\
    \textbf{Arguments:} \\
    Hidden dimension: $d$, Window size: $ws$, Epoch number: $en$, Batch size: $bs$, Learning rate: $lr$  \\
    \textbf{Output:} Training loss: $\mathcal{L}$, Model parameters: $\boldsymbol{\theta}$
    \begin{algorithmic}[1] 
    \State Get embeddings of dynamic documents $\textbf{v}_d^{'}$ from LLM
    \State Get indicator sequences $\textbf{s}_c$, $\textbf{s}_o$, $\textbf{s}_h$
    \State Split instances according to window size $ws$
    \State Sample batched training set $\mathcal{B}$ according to batch size $bs$
    \For{$epoch \in 1,...,en$}
        \For{$batch \in \mathcal{B}$}
            \State Calculate indicator features $\textbf{v}_i$ using Eq. 1-4
            \State Calculate document features $\textbf{v}_d$ using Eq. 5-6
            \State Calculate graph features $\textbf{v}_g$ using Eq. 7-9
            \State Get fused features $\textbf{H}_{i,d}^l$ from indicator and document modalities using Eq. 10-11
            \State Get stable features $\textbf{H}_{i,d}$ using Eq. 12-14
            \State Get fused features $\textbf{H}_{i,d,g}^l$ from previous features and graph modality using Eq. 15-16
            \State Get stable features $\textbf{H}_{i,d,g}$ using Eq. 17-19
            \State Get aggregated features $\textbf{h}$ using Eq. 20-22
            \State Get probability vectors $\textbf{p}$ using Eq. \ref{eq:eq9}
            \State Calculate multi-label loss $\mathcal{L}$ using Eq. \ref{eq:eq10}
            \State Return $\mathcal{L}$
        \EndFor
        \State Update $\theta$ with backpropagation from loss $\mathcal{L}$ using Adam optimizer with learning rate $lr$
    \EndFor
    \State Return model parameter $\boldsymbol{\theta}$
    \end{algorithmic}
\end{algorithm}

\section{Experiments}
We conduct experiments to answer the following research questions on the performance of MSGCA:

\begin{itemize}
    \item[\textbf{RQ1}] \textbf{Prediction performance (Section \uppercase\expandafter{\romannumeral5}-C):} Does MSGCA outperform previous methods in multimodal stock movement prediction?
    \item[\textbf{RQ2}] \textbf{Ablation study (Section \uppercase\expandafter{\romannumeral5}-D):} How do fusion strategies, language encoders, and graph encoders affect the performance of MSGCA?
    \item[\textbf{RQ3}] \textbf{Modality effect (Section \uppercase\expandafter{\romannumeral5}-E):} How do different modalities affect the performance of MSGCA? Which modality contributes the most?
    \item[\textbf{RQ4}] \textbf{Stable fusion study (Section \uppercase\expandafter{\romannumeral5}-F):} Does the gated cross-attention mechanism in MSGCA work to stabilize multimodal fusion?
    \item[\textbf{RQ5}] \textbf{Hyperparameter effect (Section \uppercase\expandafter{\romannumeral5}-G):} How do hidden dimension sizes, time window sizes and learning rates affect MSGCA performance?
    \item[\textbf{RQ6}] \textbf{Computation efficiency (Section \uppercase\expandafter{\romannumeral5}-H):} How does the training speed and memory usage of MSGCA comparing to baseline methods?
\end{itemize}

\subsection{Experimental Setup}
\textit{1) Datasets: } We evaluate the performance of MSGCA on three public benchmark datasets: \textbf{BigData22} \cite{soun2022accurate}, \textbf{ACL18} \cite{xu2018stock}, \textbf{CIKM18} \cite{wu2018hybrid}, and one new dataset that we collect and publish: \textbf{InnoStock}. Table \ref{dataset} presents a summary of these datasets, detailing the count of stocks, the number of news documents (InnoStock) and tweet documents (others), the number of edges in the industry graph and the duration of trading dates. The three existing datasets consist of high-trade-volume stocks in US stock markets and dynamic input from tweets and price sequences. Innostock, originally collected by us from CSMAR \footnote{https://cn.gtadata.com/}, focuses on newly formed technology companies listed on China's Sci-Tech Innovation Board, aggregating their financial news from various online platforms. To accommodate MSGCA's multimodal input, we further collect the industrial sector relationships for each stock from the above datasets to build graphs. We label each stock according to the increase rate of its adjusted closed prices. Following \cite{soun2022accurate}, the increase rate is calculated as $r_t^s = p_t^s / p_{t-1}^s - 1$, where $p_t^i$ is the adjusted closing price of the stock $s$ at the timestamp $t$. To ensure balanced labels in our fine-grained classification, we adjust the growth rate threshold for flat trends according to each dataset. We set $[-1\%, 1\%]$, $[-0.5\%, 0.5\%]$, $[-0.4\%, 0.4\%]$, and $[-0.3\%, 0.3\%]$ for InnoStock, BigData22, ACL18, and CIKM18, respectively. For example, InnoStock timestamps that have $r_t^s \geq 1\%$ and $r_t^s \leq -1\%$ are labeled as up (1) and down (-1), respectively, and $-1\% < r_t^s < 1\%$ are labeled as flat (0). We chronologically partitioned each dataset into training, validation, and testing subsets, consistent with recent studies on stock movement prediction \cite{soun2022accurate} \cite{yoo2021accurate}.

\begin{table}[htbp]
\caption{Summary of Datasets Statistics.}
\centering
\begin{tabular}{l|c|c|c|c}
\bottomrule
\textbf{Datasets} & \textbf{Stocks} & \textbf{Documents} & \textbf{Edges} & \textbf{Dates}
\\ \hline   
InnoStock & 369 & 6,756 & 385 & 2022-01-04 to 2022-12-30
\\
BigData22 & 50 & 272,762 & 50 & 2019-07-05 to 2020-06-30
\\
CIKM18 & 38 & 955,788 & 26 & 2017-01-03 to 2017-12-28
\\
ACL18 & 87 & 106,271 & 87 & 2014-01-02 to 2015-12-30
\\ \bottomrule
\end{tabular}
\label{dataset}
\end{table}

\textit{2) Evaluation Metrics: } We assess the performance of multimodal stock movement prediction using two measures: accuracy (\textbf{ACC}) and the Matthews Correlation Coefficient (\textbf{MCC}) \cite{matthews1975comparison}. ACC is commonly used in various classification issues, while MCC improves the fairness of the evaluation by considering all four confusion matrix indicators: true positives ($tp$), true negatives ($tn$), false positives ($fp$) and false negatives ($fn$). The MCC can be calculated as follows:

\begin{equation}
\begin{aligned}
MCC = \frac{tp \times tn - fp \times fn}{\sqrt{(tp+fp)(tp+fn)(tn+fp)(tn+fn)}}.
\label{eq:eq12}
\end{aligned}
\end{equation}

\textit{3) Hyperparameters: } We train our MSGCA and all baseline methods with the following key hyperparameters. The hidden embedding dimensions of all the encoder output and fused stock features are set to 64. The size of the sequence window is set to 20. We set the number of heads to 2 for multi-head cross-attention layers. The number of training epochs is 200 for BigData22 and CIKM18 and 250 for InnoStock and ACL18. The batch size is set to 4096 for InnoStock and 1024 for others. The learning rate is 1e-4. We use a warm-up training strategy for all methods with the Adam optimizer. All hyperparameters of the baseline methods are set as the numbers reported in the original papers. We run each method five times with different random seeds and report the average performance and variances.

\subsection{Baseline Methods}
We compare the performance of our MSGCA with three categories of baselines for multimodal stock movement prediction as follows: 

1) \textit{Indicator-only methods}, which focus only on price input to perform prediction, including \textbf{LSTM} \cite{nelson2017stock} and \textbf{ALSTM} \cite{qin2017dual}. LSTM is a representative model for sequential data. ALSTM combines the hidden states of LSTM with the output of attention.

2) \textit{Indicator-document methods}, which combine the features of indicator sequences and dynamic documents to predict future trends, including \textbf{ALSTM-W} and \textbf{SLOT} \cite{soun2022accurate}. ALSTM-W computes the average embeddings of documents using a pre-trained language model along with indicator features. SLOT integrates price features and two types of trend features with ALSTM. 

3) \textit{Indicator-graph methods}, that use graph structural information to improve performance, including \textbf{ESTIMATE} \cite{huynh2023efficient} and \textbf{DTML} \cite{yoo2021accurate}. ESTIMATE concatenates price features with stock hypergraph features for prediction. DTML uses attentive LSTM to compute features of stocks and market relations and integrates them with concatenation and attention.

We run the above baseline methods by either modifying their open-source code to adjust our datasets and task, or reimplementing referring to their original papers. 

\subsection{Performance Comparison (RQ1)}
Stock movement prediction performance are evaluated with the two metrics. The results are shown in Table \ref{performance}, where the best results are highlighted in bold, and the most competitive baselines are underlined. Our MSGCA consistently achieves the best performance on all datasets and metrics. Specifically, we have the following observations.
\begin{table*}[htbp]
\caption{Results of Stock Movement Prediction on Four Datasets and Two Metrics.}
\centering
\begin{tabular}{l|cccccccc}
\bottomrule
\multirow{2}{*}{\textbf{Methods}} & \multicolumn{2}{c}{\textbf{InnoStock}} & \multicolumn{2}{c}{\textbf{BigData22}} & \multicolumn{2}{c}{\textbf{ACL18}} & \multicolumn{2}{c}{\textbf{CIKM18}}
\\ &
\multicolumn{1}{c}{\textbf{ACC}} & \multicolumn{1}{c}{\textbf{MCC}} & \multicolumn{1}{c}{\textbf{ACC}} & \multicolumn{1}{c}{\textbf{MCC}} & \multicolumn{1}{c}{\textbf{ACC}} & \multicolumn{1}{c}{\textbf{MCC}} & \multicolumn{1}{c}{\textbf{ACC}} & \multicolumn{1}{c}{\textbf{MCC}}
\\ \hline      
\multirow{1}{*}{LSTM} & \multicolumn{1}{c}{0.3545{\scriptsize±0.0005}} & \multicolumn{1}{c}{0.0320{\scriptsize±0.0011}} & \multicolumn{1}{c}{0.4119{\scriptsize±0.0023}} & \multicolumn{1}{c}{0.0866{\scriptsize±0.0059}} & \multicolumn{1}{c}{0.3779{\scriptsize±0.0022}} & \multicolumn{1}{c}{0.0254{\scriptsize±0.0020}} & \multicolumn{1}{c}{0.3781{\scriptsize±0.0006}} & \multicolumn{1}{c}{0.0393{\scriptsize±0.0028}}
\\
\multirow{1}{*}{ALSTM} & \multicolumn{1}{c}{0.3762{\scriptsize±0.0017}} & \multicolumn{1}{c}{0.0339{\scriptsize±0.0014}} & \multicolumn{1}{c}{0.4335{\scriptsize±0.0079}} & \multicolumn{1}{c}{0.0968{\scriptsize±0.0072}} & \multicolumn{1}{c}{0.3806{\scriptsize±0.0027}} & \multicolumn{1}{c}{0.0275{\scriptsize±0.0031}} & \multicolumn{1}{c}{0.3830{\scriptsize±0.0011}} & \multicolumn{1}{c}{0.0470{\scriptsize±0.0046}}
\\ 
\multirow{1}{*}{DTML} & \multicolumn{1}{c}{\underline{0.4208{\scriptsize±0.0009}}} & \multicolumn{1}{c}{0.0493{\scriptsize±0.0013}} & \multicolumn{1}{c}{\underline{0.4367{\scriptsize±0.0123}}} & \multicolumn{1}{c}{\underline{0.1048{\scriptsize±0.0131}}} & \multicolumn{1}{c}{0.3832{\scriptsize±0.0053}} & \multicolumn{1}{c}{0.0306{\scriptsize±0.0085}} & \multicolumn{1}{c}{0.3811{\scriptsize±0.0080}} & \multicolumn{1}{c}{\underline{0.0613{\scriptsize±0.0072}}}
\\
\multirow{1}{*}{ESTIMATE} & \multicolumn{1}{c}{0.3995{\scriptsize±0.0020}} & \multicolumn{1}{c}{0.4820{\scriptsize±0.0046}} & \multicolumn{1}{c}{0.4296{\scriptsize±0.0027}} & \multicolumn{1}{c}{0.1004{\scriptsize±0.0059}} & \multicolumn{1}{c}{0.3805{\scriptsize±0.0046}} & \multicolumn{1}{c}{0.0312{\scriptsize±0.0074}} & \multicolumn{1}{c}{0.3806{\scriptsize±0.0047}} & \multicolumn{1}{c}{0.0579{\scriptsize±0.0035}}
\\ 
\multirow{1}{*}{ALSTM-W} & \multicolumn{1}{c}{0.3835{\scriptsize±0.0015}} & \multicolumn{1}{c}{0.0431{\scriptsize±0.0022}} & \multicolumn{1}{c}{0.4287{\scriptsize±0.0025}} & \multicolumn{1}{c}{0.0999{\scriptsize±0.0066}} & \multicolumn{1}{c}{0.3848{\scriptsize±0.0050}} & \multicolumn{1}{c}{0.0375{\scriptsize±0.0064}} & \multicolumn{1}{c}{0.3818{\scriptsize±0.0082}} & \multicolumn{1}{c}{0.0545{\scriptsize±0.0096}}
\\
\multirow{1}{*}{SLOT} & \multicolumn{1}{c}{0.4127{\scriptsize±0.0006}} & \multicolumn{1}{c}{\underline{0.0509{\scriptsize±0.0005}}} & \multicolumn{1}{c}{0.4250{\scriptsize±0.0082}} & \multicolumn{1}{c}{0.0980{\scriptsize±0.0081}} & \multicolumn{1}{c}{\underline{0.3894{\scriptsize±0.0016}}} & \multicolumn{1}{c}{\underline{0.0487{\scriptsize±0.0168}}} & \multicolumn{1}{c}{\underline{0.3848{\scriptsize±0.0011}}} & \multicolumn{1}{c}{0.0601{\scriptsize±0.0053}}
\\ \hline
\multirow{1}{*}{MSGCA} & \multicolumn{1}{c}{\textbf{0.4223{\scriptsize±0.0003}}} & \multicolumn{1}{c}{\textbf{0.0550{\scriptsize±0.0008}}} & \multicolumn{1}{c}{\textbf{0.4379{\scriptsize±0.0077}}} & \multicolumn{1}{c}{\textbf{0.1112{\scriptsize±0.0037}}} & \multicolumn{1}{c}{\textbf{0.3966{\scriptsize±0.0050}}} & \multicolumn{1}{c}{\textbf{0.0593{\scriptsize±0.0073}}} & \multicolumn{1}{c}{\textbf{0.3861{\scriptsize±0.0023}}} & \multicolumn{1}{c}{\textbf{0.0807{\scriptsize±0.0045}}}
\\ \bottomrule
\end{tabular}
\label{performance}
\end{table*}

\textbf{\textit{1) MSGCA outperforms baselines on all datasets:}} Due to the introduction of more modalities and the stable fusion modules, our MSGCA achieves the best results on all datasets and metrics for the fine-grained stock movement prediction task. Considering the MCC metric, the improvements in MSGCA compared to the second-best results on four datasets are 8.1\%, 6.1\%, 21.7\% and 31.6\%, respectively. Meanwhile, the best competitors on four datasets are SLOT or DTML. These two methods have been proposed in recent years and have more advanced architectures than other baselines, which shows consistency with their results.

\textbf{\textit{2) Modalities contribute differently on each dataset:}} Given the variability in stocks, time spans, and data origins, the three modalities could have varying impacts on predictive accuracy. Regarding the performance of baseline methods, dynamic documents of stocks has a greater impact on ACL18 and InnoStock compared to its influence on BigData22 and CIKM18. On the other hand, industry relationships, which facilitate the propagation of information between stocks, prove to be more effective in BigData22 compared to others. However, the trimodal encoder and gated cross-attention of our MSGCA efficiently integrate information across all three modalities, enhancing overall performance.

\textbf{\textit{3) Introducing more modalities may damage accuracy:}} Inconsistent multimodal datasets cause unstable feature fusion in baseline methods, leading to inaccurate predictions. Specifically, when integrating dynamic document modalities, we observe a relative decline in accuracy on BigData22, ACL18, and CIKM18 datasets compared to the unimodal ALSTM frameworks. However, our MSGCA effectively processes noisy document data and achieves accurate predictions, leveraging a gated cross-attention mechanism for reliable integration. Additionally, our InnoStock dataset reveals a consistent performance improvement when integrating news documents and relational graphs, demonstrating the superior quality of our proposed datasets.

\subsection{Ablation Study (RQ2)}
In this subsection, we examine how various components within the encoding and fusion modules impact MSGCA's performance. We denote the MSGCA variants as \textquote{MSGCA-X}, which represents the framework to replace a specific part of it with \textquote{X}. Our study focuses on three aspects as follows.

\begin{figure}[htbp]
\centerline{\includegraphics[width=0.5\textwidth]{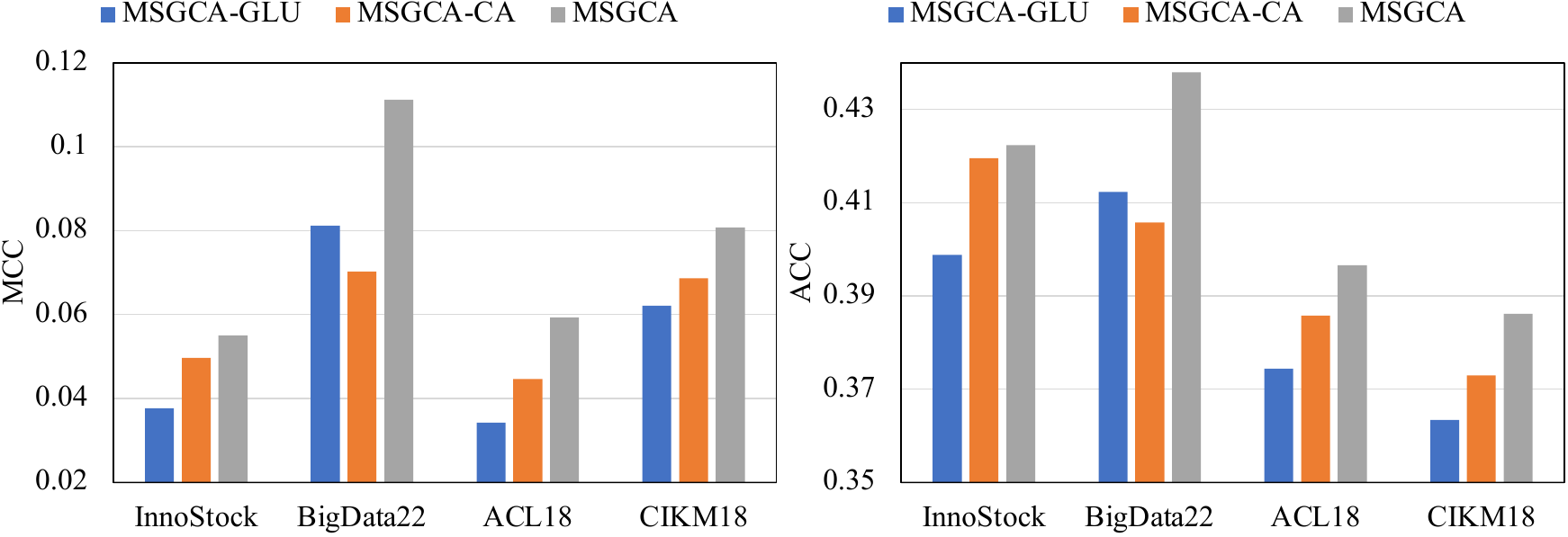}}
\caption{Performance impacts of different fusion strategies on four datasets. MSGCA-GLU cannot manage the interaction between multimodal features, while MSGCA-CA suffers from noisy information aggregation.}
\label{ablation1}
\end{figure}

\textbf{\textit{1) Impact of different fusion strategies: }} We evaluate the impact of fusion strategies by replacing gated cross-attention blocks in MSGCA with two foundational networks: gated layer unit (GLU) \cite{dauphin2017language} and cross-attention \cite{rajan2022cross}. These two variants are denoted as \textbf{MSGCA-GLU} and \textbf{MSGCA-CA}, respectively. From the performance comparison of the three frameworks shown in Fig. \ref{ablation1}, the GLU fusion approach is direct yet overly simplistic to effectively manage the interactions between multimodal features for competitive predictions. The cross-attention mechanism can effectively aggregate features from multiple modalities, but suffers from noisy information and leads to performance decline. Furthermore, the results on BigData22 indicate a more detrimental fusion between modalities compared to other datasets.

\begin{figure}[htbp]
\centerline{\includegraphics[width=0.5\textwidth]{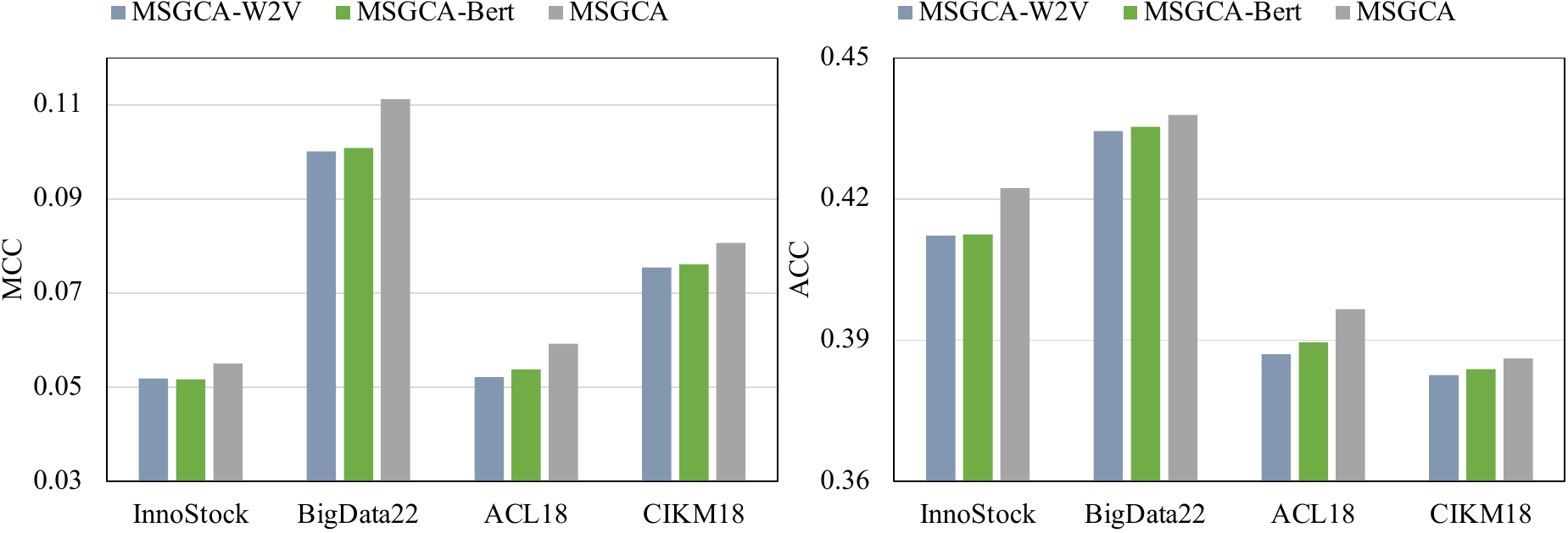}}
\caption{Impacts of different language encoding models on four datasets for handling dynamic documents. Embeddings generated from the advanced large language model can significantly improve the performance.}
\label{ablation2}
\end{figure}

\textbf{\textit{2) Impact of language encoding models: }} A language encoding model transforms document text into latent embeddings for the subsequent fusion process. In our study, we evaluated three pretrained language models, including Word2Vec \cite{mikolov2013efficient}, BERT \cite{devlin2018bert}, and Ada \footnote{https://openai.com/blog/new-and-improved-embedding-model} to preprocess and manage dynamic document modality input. These language models vary in parameter size and output dimensionality. The frameworks using these three language models are denoted as \textbf{MSGCA-W2V}, \textbf{MSGCA-Bert}, and \textbf{MSGCA}, respectively. The results shown in Fig. \ref{ablation2} illustrate that prediction performance improves along with the size of the language model employed. Furthermore, the advanced large language model (Ada) used into our final version of MSGCA outperforms the two smaller language models, delivering superior results.

\begin{figure}[htbp]
\centerline{\includegraphics[width=0.5\textwidth]{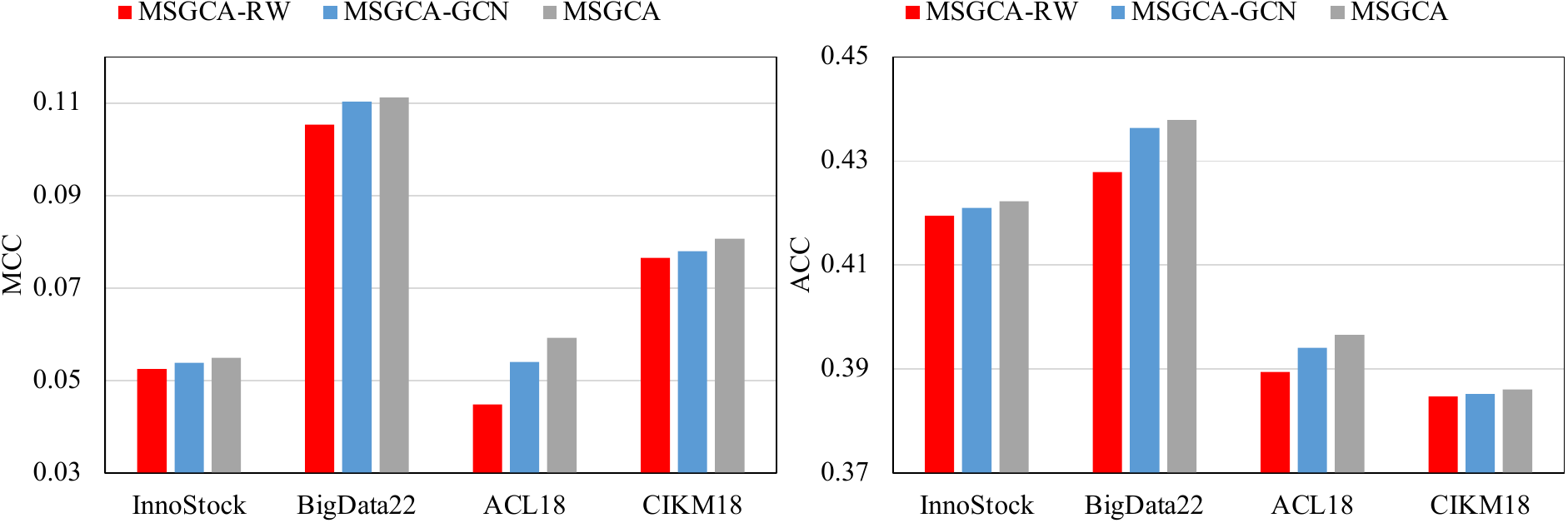}}
\caption{Impacts of different graph encoding models on four datasets for handling graph structural information. Graph attention network  enables nodes to attend over their neighboring features and helps the most for the overall performance.}
\label{ablation3}
\end{figure}

\textbf{\textit{3) Impact of graph encoding models: }} To evaluate the impact of graph encoding models, we extract structural insights from stock interrelations by applying multiple graph encoding techniques, such as Random Walk \cite{perozzi2014deepwalk}, GCN \cite{kipf2016semi}, and GAT \cite{velivckovic2017graph}. These models aggregate information from neighboring nodes with different selection and computation strategies to obtain central node features. We denote frameworks using these three graph encoders as \textbf{MSGCA-RW}, \textbf{MSGCA-GCN}, and \textbf{MSGCA}. As shown in Fig. \ref{ablation3}, models considering a higher order of relationships can achieve better performance. Our MSGCA, which employs GAT, can implicitly specify different weights to all nodes in a neighborhood while encoding graph structures and finally gains the best results compared to other graph encoding models.

\subsection{Effect of Multimodal Sources (RQ3)}
In this subsection, we examine how various modalities aid in predicting stock movement. To evaluate this, we implement three variants of MSGCA where a modality is removed from our trimodal encoder module: 1) \textbf{MSGCA-ID}: a version in which indicator sequences and dynamic documents are considered without a relational graph as input; 2) \textbf{MSGCA-IG}: a version that indicator sequences and graph are kept for feature fusion; 3) \textbf{MSGCA-DG}: a version using only documents and relationships, excluding the primary indicator modality. The performance of these variants on four datasets is shown in Table \ref{modalsource}. We can observe from the results that MSGCA-ID and MSGCA-IG still outperform existing stock movement prediction methods on most metrics, demonstrating the capability of our feature fusion and movement prediction modules. For example, the MCC of MSGCA-ID is 1.7\% and 9.8\% higher than that of SLOT on InnoStock and BigData22, respectively. In addition, the modality of indicator sequences plays a crucial role in the prediction performance. All results of MSGCA-DG decrease dramatically after the indicator modality is removed from those of other frameworks. Furthermore, without relying on indicator features, MSGCA can yield results compared to baseline methods that use indicators alone, showing its capability to effectively utilize the information from graph relationships and the textual data of news and tweets for prediction.

\begin{table*}[htbp]
\caption{Effects of Different Modalities on Stock Movement Prediction.}
\centering
\begin{tabular}{l|cccccccc}
\bottomrule
\multirow{2}{*}{\textbf{Methods}} & \multicolumn{2}{c}{\textbf{InnoStock}} & \multicolumn{2}{c}{\textbf{BigData22}} & \multicolumn{2}{c}{\textbf{ACL18}} & \multicolumn{2}{c}{\textbf{CIKM18}}
\\ &
\multicolumn{1}{c}{\textbf{ACC}} & \multicolumn{1}{c}{\textbf{MCC}} & \multicolumn{1}{c}{\textbf{ACC}} & \multicolumn{1}{c}{\textbf{MCC}} & \multicolumn{1}{c}{\textbf{ACC}} & \multicolumn{1}{c}{\textbf{MCC}} & \multicolumn{1}{c}{\textbf{ACC}} & \multicolumn{1}{c}{\textbf{MCC}}
\\ \hline      
\multirow{1}{*}{MSGCA-ID} & \multicolumn{1}{c}{0.4163} & \multicolumn{1}{c}{0.0518} & \multicolumn{1}{c}{0.4309} & \multicolumn{1}{c}{0.1076} & \multicolumn{1}{c}{0.3860} & \multicolumn{1}{c}{0.0354} & \multicolumn{1}{c}{0.3811} & \multicolumn{1}{c}{0.0664}
\\
\multirow{1}{*}{MSGCA-IG} & \multicolumn{1}{c}{0.4163} & \multicolumn{1}{c}{0.0525} & \multicolumn{1}{c}{0.4273} & \multicolumn{1}{c}{0.1006} & \multicolumn{1}{c}{0.3889} & \multicolumn{1}{c}{0.0499} & \multicolumn{1}{c}{0.3835} & \multicolumn{1}{c}{0.0738}
\\ 
\multirow{1}{*}{MSGCA-DG} & \multicolumn{1}{c}{0.3569} & \multicolumn{1}{c}{0.0344} & \multicolumn{1}{c}{0.4052} & \multicolumn{1}{c}{0.0775} & \multicolumn{1}{c}{0.3591} & \multicolumn{1}{c}{0.0189} & \multicolumn{1}{c}{0.3778} & \multicolumn{1}{c}{0.0401}
\\
\multirow{1}{*}{MSGCA} & \multicolumn{1}{c}{\textbf{0.4223}} & \multicolumn{1}{c}{\textbf{0.0550}} & \multicolumn{1}{c}{\textbf{0.4379}} & \multicolumn{1}{c}{\textbf{0.1112}} & \multicolumn{1}{c}{\textbf{0.3966}} & \multicolumn{1}{c}{\textbf{0.0593}} & \multicolumn{1}{c}{\textbf{0.3861}} & \multicolumn{1}{c}{\textbf{0.0807}}
\\ \bottomrule
\end{tabular}
\label{modalsource}
\end{table*}

\subsection{Study for Stable Fusion (RQ4)}
\begin{figure}[htbp]
\centerline{\includegraphics[width=0.5\textwidth]{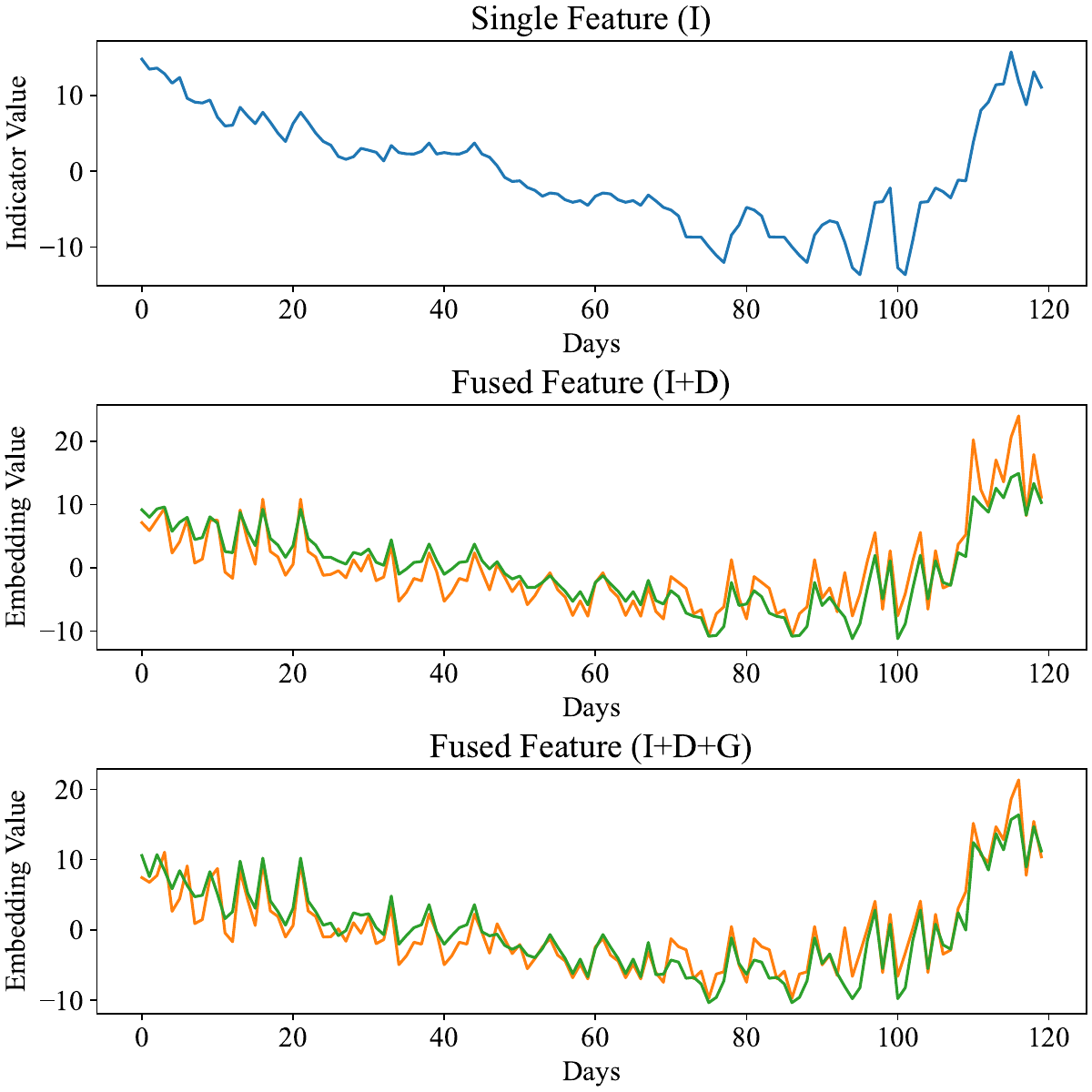}}
\caption{Stable fusion study by reducing embedding dimensions of a stock to a single dimension, allowing for direct comparison with its price trend. Blue lines represent the price time-series of the stock. Yellow lines are features after cross-attention fusion for aggregating document and graph modalities without guidance by the primary modality. Green lines are features from gated cross-attention blocks, exhibiting smoother transitions than yellow lines and indicating stable fusion effects.}
\label{stable}
\end{figure}

To demonstrate that our MSGCA is able to perform a stable feature fusion for multimodal stock movement prediction, we further analyze the reliability of MSGCA during the progressive fusion phase across three modalities. We randomly select a representative stock from the InnoStock dataset. Embedding vectors of the stock, derived from the fused features, are sequentially dumped to files following each of the two fusion stages. Then, we reduce the feature dimensions into one-dimensional vectors via Principal Component Analysis (PCA), which are then aligned with the stock's closing price timeline for comparative analysis across various fusion stages. This comparison is shown in Fig. \ref{stable}. Visualizing the latent features of InnoStock reveals that the gated layers of MSGCA, coupled with a cross-attention mechanism, can mitigate issues of data sparsity and semantic conflicts. To be more specific, two green lines, produced by gated cross-attention blocks, are smoother and more aligned with the price trajectory compared to the yellow lines, which are derived from feature fusion without primary modality guidance and display significant volatility. Meanwhile, the smoothing effect takes place after each fusion stage (I+D and I+D+G), indicating that our gated cross-attention fusion strategy consistently addresses the data challenge across multiple modalities.

\subsection{Effect of Hyperparameters (RQ5)}
\begin{figure}[htbp]
\centerline{\includegraphics[width=0.5\textwidth]{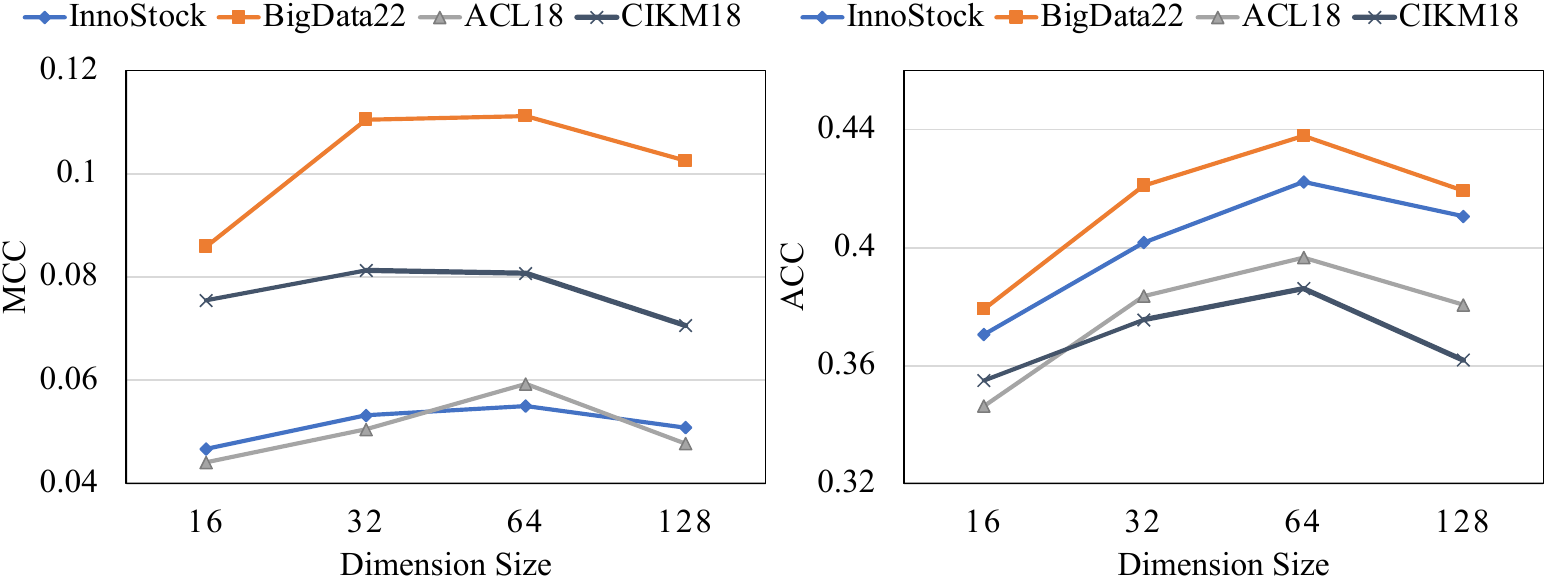}}
\caption{The MCC of MSGCA with various dimension sizes of hidden embeddings on four datasets.}
\label{hyperparam1}
\end{figure}

\textit{1) Impact of hidden dimension size: } We investigate the impact of different dimension sizes of hidden embeddings in MSGCA, including the representations of each modality and the features after fusion. We set the dimension size $d \in [16, 32, 64, 128]$. The MCC and ACC variation curves for the four datasets are shown in Fig. \ref{hyperparam1}. We can observe that the performance of MSGCA increases along with the sizes from 16 to 64, and decreases with the size of 128 on all datasets. This illustrates a lack of expression with smaller dimensions and an overfitting with larger dimensions. In addition, ACL18 and BigData22 show more sensitive to hidden dimension size than the other two datasets. Considering both the performance and efficiency, we choose $d=64$ as the best hidden dimension size.

\begin{figure}[htbp]
\centerline{\includegraphics[width=0.5\textwidth]{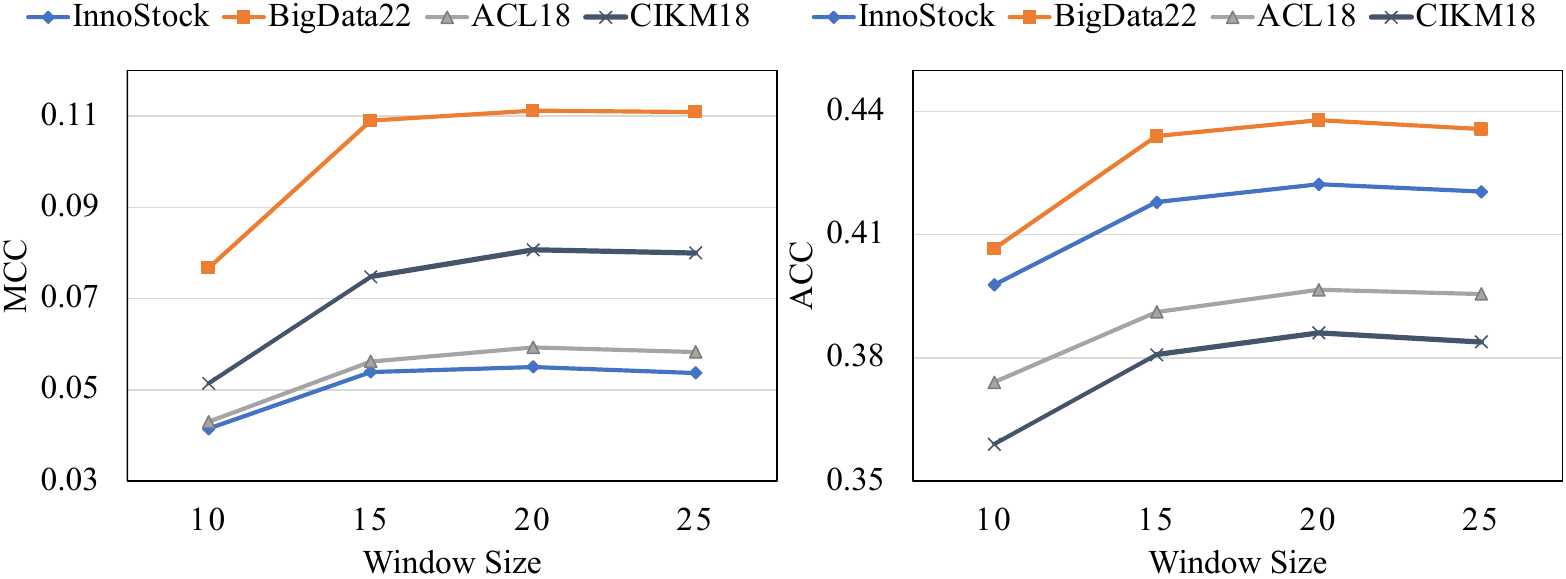}}
\caption{The Performance of MSGCA with various window sizes for input sequences on four datasets.}
\label{hyperparam2}
\end{figure}

\textit{2) Impact of time window size: } We evaluate the influence of window size in MSGCA. Following settings from previous studies \cite{yoo2021accurate} \cite{soun2022accurate}, we set the window size $ws \in [10, 15, 20, 25]$ which balances the length of sequences and the amount of training samples. The MCC and ACC results on four datasets with different window sizes are shown in Fig. \ref{hyperparam2}. The performance of MSGCA dramatically decreases with window size of 10 due to a lack of useful information from the time series. Meanwhile, MSGCA cannot gain further improvement from larger window sizes because of fewer training samples obtained from datasets, which limits the information to be learned. In our experiments with the best results, we set $ws = 20$ as the window size.

\begin{figure}[htbp]
\centerline{\includegraphics[width=0.5\textwidth]{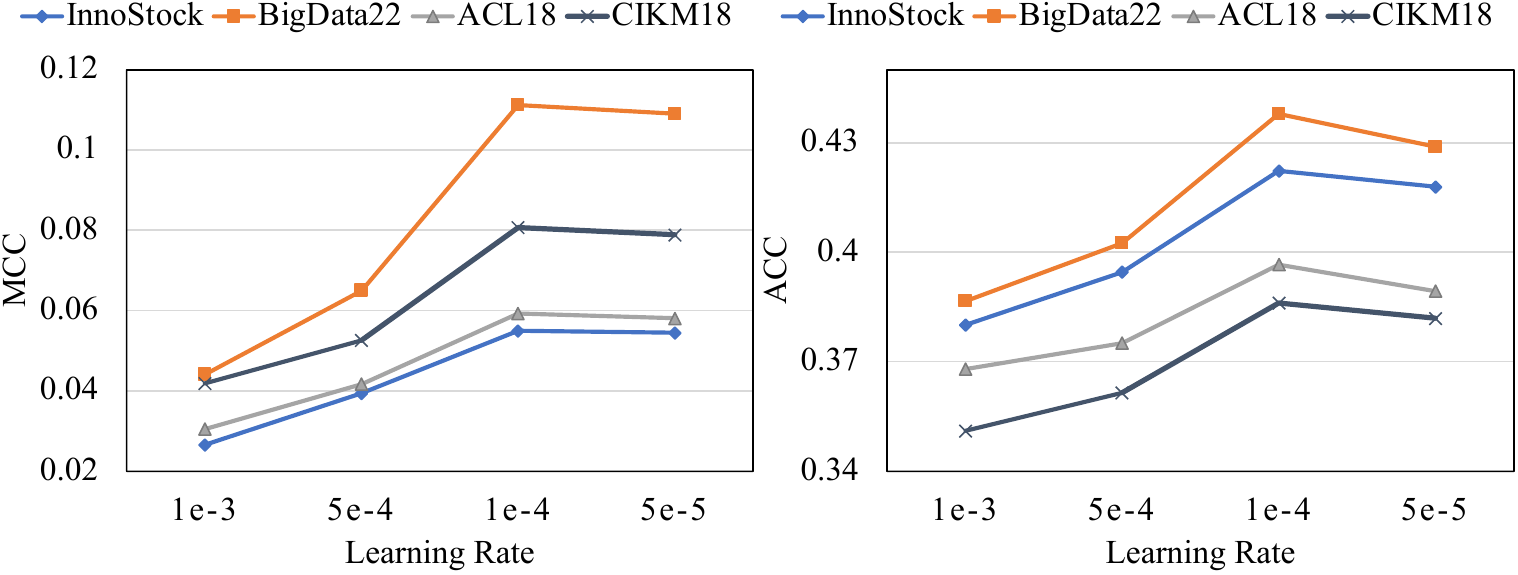}}
\caption{Performance of MSGCA with various learning rates on four datasets.}
\label{hyperparam3}
\end{figure}

\textit{3) Impact of learning rate: } Training strategies significantly affect the performance of MSGCA, and the learning rate is a crucial training parameter. As shown in Fig. \ref{hyperparam3}, a too slow gradient descent (with the learning rate set to $5e-5$) can hardly reach optimal results within reasonable training steps, leading to a low-efficiency training process. Meanwhile, a training speed that is too fast (with the learning rate set to $1e-3$ and $5e-4$) may skip optimal results and generate suboptimal performance. With a large number of experiments, we set the learning rate $lr = 1e-4$ as the best in MSGCA.

\subsection{Computation Efficiency Analysis (RQ6)}
\begin{figure}[htbp]
\centerline{\includegraphics[width=0.5\textwidth]{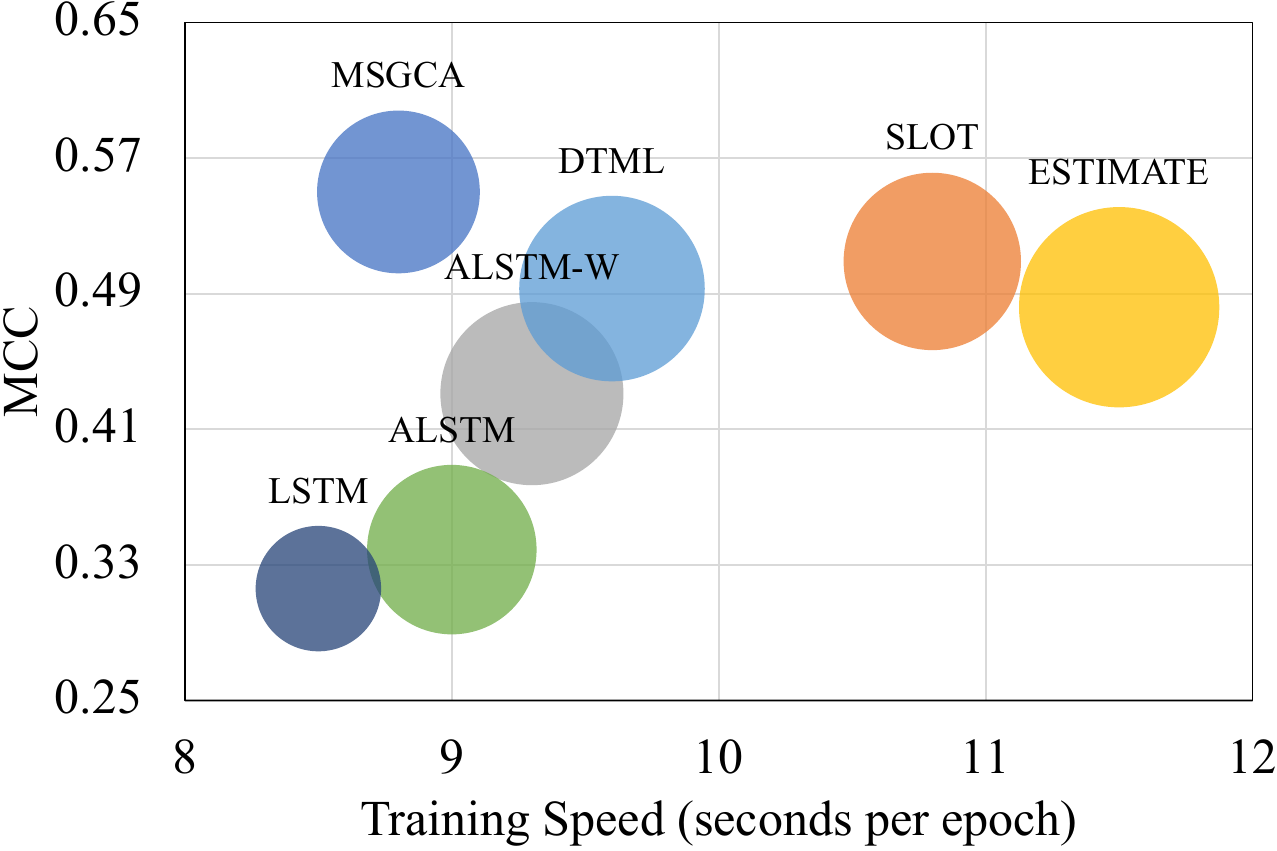}}
\caption{Computation efficiency comparison on InnoStock dataset. MCC performance ($y$ axis), training speed ($x$ axis), and memory footprint (size of the circles) of methods are shown.}
\label{cost}
\end{figure}

In this subsection, we analyze the computation efficiency of MSGCA comparing with baseline methods in two aspects, including the training speed and the memory cost. We evaluate all methods with the same batch size and record the average time usage per epoch for each method. On the other hand, we monitor the CUDA memory usage by dumping snapshots of memory allocations during training and record the largest memory value for each method. The comparison of computation efficiency on InnoStock dataset is shown in Fig. \ref{cost}. For one aspect, our MSGCA delivers superior results with a faster training process and smaller memory footprint than current leading stock movement prediction methods. For another, MSGCA achieves cost-effective computation while significantly outperforms methods with simpler architecture.

\section{Conclusion and Future Work}
In our study, we examine the challenges of multimodal stock movement prediction using a stable fusion approach via a gated cross-attention mechanism. Our contribution, MSGCA, addresses the task by leveraging heterogeneous data sources and fills the gap with the study of the fine-grained stock movement prediction task with multimodal information. The MSGCA framework begins with a trimodal encoder to handle three distinct data modalities: indicator sequences, dynamic documents, and a relational graph. This is followed by a stable fusion module to subsequently integrate these three modalities with a pair of gated cross-attention networks guided by primary features. The final phase involves deploying a movement prediction module consisting of two MLP networks to map the integrated features into predictive signals to perform the fine-grained prediction task. Extensive experiments and analysis on four multimodal stock datasets demonstrate the effectiveness and efficiency of MSGCA. In future work, we aim to enrich our model with more features from various modalities and explore its fusion capabilities in different application domains.

\section*{Acknowledgment}
We would like to thank the anonymous reviewer for his/her careful reading of our manuscript and his/her many insightful comments and suggestions.

\begin{IEEEbiography}[{\includegraphics[width=1in,height=1.25in,clip,keepaspectratio]{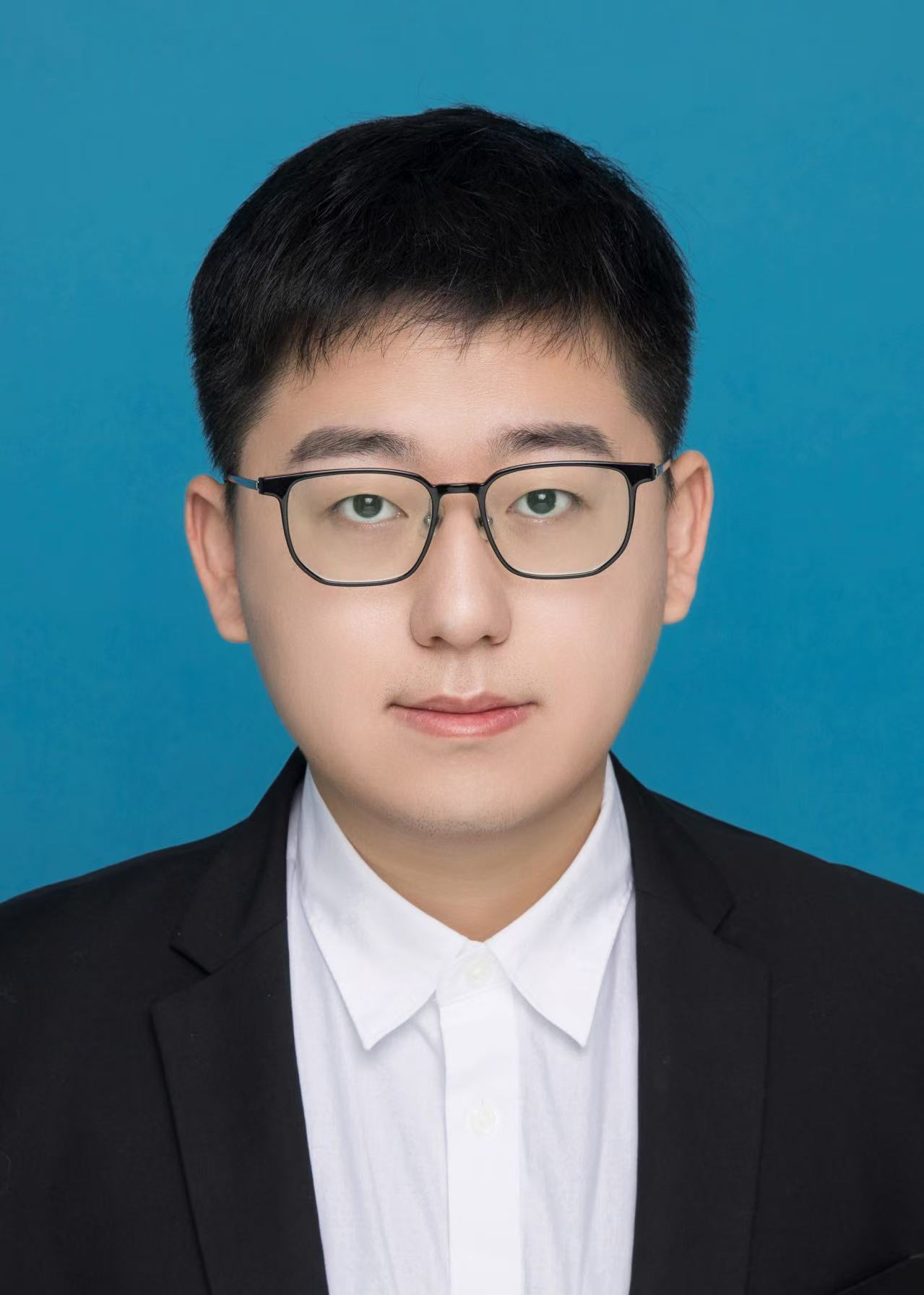}}]{Chang Zong}
Chang Zong received the doctoral degree from Zhejiang University, Hangzhou, China. He is currently working as a researcher at Zhejiang University of Science and Technology. His research interests include natural language processing, knowledge graph, and data mining applications in finance and medicine. \end{IEEEbiography}

\begin{IEEEbiography}[{\includegraphics[width=1in,height=1.25in,clip,keepaspectratio]{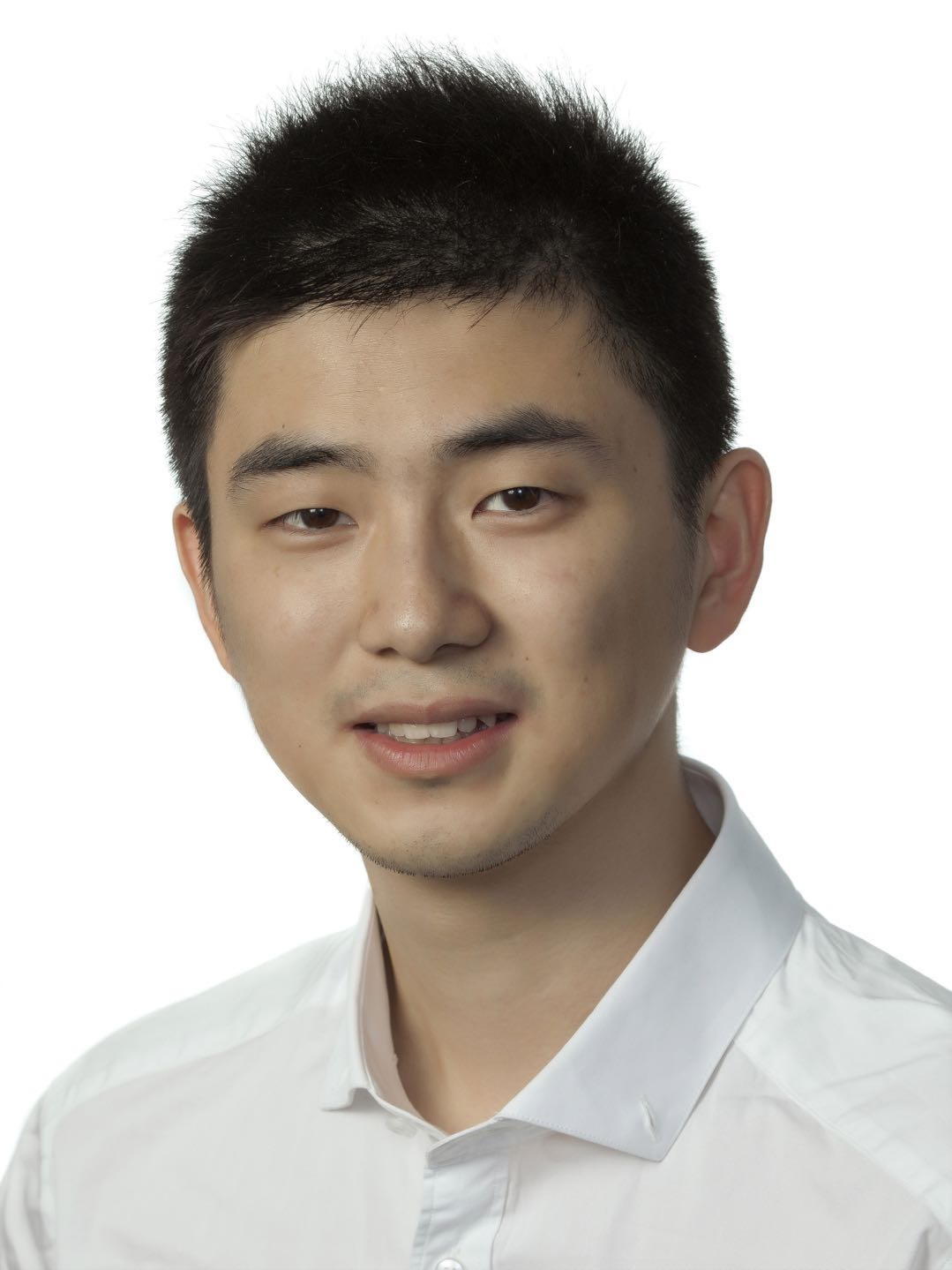}}]{Zhou Hang}
Hang Zhou received the doctoral degree from the University of Edinburgh. He is currently working as an assistant professor in finance at the Nottingham University Business School China. His research is mainly in the areas of capital market asset pricing, credit risk, and machine learning applications. \end{IEEEbiography}

\vfill


\begin{thebibliography}{00}

\bibitem{feng2019temporal}
F.~Feng, X.~He, X.~Wang, C.~Luo, Y.~Liu, and T.-S. Chua, ``Temporal relational ranking for stock prediction,'' \emph{ACM Transactions on Information Systems (TOIS)}, vol.~37, no.~2, pp. 1--30, 2019.

\bibitem{nelson2017stock}
D.~M. Nelson, A.~C. Pereira, and R.~A. De~Oliveira, ``Stock market's price movement prediction with lstm neural networks,'' in \emph{2017 International joint conference on neural networks (IJCNN)}.\hskip 1em plus 0.5em minus 0.4em\relax Ieee, 2017, pp. 1419--1426.

\bibitem{long2019deep}
W.~Long, Z.~Lu, and L.~Cui, ``Deep learning-based feature engineering for stock price movement prediction,'' \emph{Knowledge-Based Systems}, vol. 164, pp. 163--173, 2019.

\bibitem{nguyen2015sentiment}
T.~H. Nguyen, K.~Shirai, and J.~Velcin, ``Sentiment analysis on social media for stock movement prediction,'' \emph{Expert Systems with Applications}, vol.~42, no.~24, pp. 9603--9611, 2015.

\bibitem{liu2019transformer}
J.~Liu, H.~Lin, X.~Liu, B.~Xu, Y.~Ren, Y.~Diao, and L.~Yang, ``Transformer-based capsule network for stock movement prediction,'' in \emph{Proceedings of the first workshop on financial technology and natural language processing}, 2019, pp. 66--73.

\bibitem{gao2021graph}
J.~Gao, X.~Ying, C.~Xu, J.~Wang, S.~Zhang, and Z.~Li, ``Graph-based stock recommendation by time-aware relational attention network,'' \emph{ACM Transactions on Knowledge Discovery from Data (TKDD)}, vol.~16, no.~1, pp. 1--21, 2021.

\bibitem{xu2018stock}
Y.~Xu and S.~B. Cohen, ``Stock movement prediction from tweets and historical prices,'' in \emph{Proceedings of the 56th Annual Meeting of the Association for Computational Linguistics (Volume 1: Long Papers)}, 2018, pp. 1970--1979.

\bibitem{ma2023multi}
Ma, Y., Mao, R., Lin, Q., Wu, P. \& Cambria, E. Multi-source aggregated classification for stock price movement prediction. {\em Information Fusion}. 91 pp. 515-528 (2023)

\bibitem{sawhney2020deep}
R.~Sawhney, S.~Agarwal, A.~Wadhwa, and R.~Shah, ``Deep attentive learning for stock movement prediction from social media text and company correlations,'' in \emph{Proceedings of the 2020 Conference on Empirical Methods in Natural Language Processing (EMNLP)}, 2020, pp. 8415--8426.

\bibitem{li2021modeling}
Li, W., Bao, R., Harimoto, K., Chen, D., Xu, J. \& Su, Q. Modeling the stock relation with graph network for overnight stock movement prediction. {\em Proceedings Of The Twenty-ninth International Conference On International Joint Conferences On Artificial Intelligence}. pp. 4541-4547 (2021)

\bibitem{xie2023wall}
Q.~Xie, W.~Han, Y.~Lai, M.~Peng, and J.~Huang, ``The wall street neophyte: A zero-shot analysis of chatgpt over multimodal stock movement prediction challenges,'' \emph{arXiv preprint arXiv:2304.05351}, 2023.

\bibitem{kaeley2023support}
H.~Kaeley, Y.~Qiao, and N.~Bagherzadeh, ``Support for stock trend prediction using transformers and sentiment analysis,'' \emph{arXiv preprint arXiv:2305.14368}, 2023.

\bibitem{soun2022accurate}
Y.~Soun, J.~Yoo, M.~Cho, J.~Jeon, and U.~Kang, ``Accurate stock movement prediction with self-supervised learning from sparse noisy tweets,'' in \emph{2022 IEEE International Conference on Big Data (Big Data)}.\hskip 1em plus 0.5em minus 0.4em\relax IEEE, 2022, pp. 1691--1700.

\bibitem{zhao2017time}
Zhao, Z., Rao, R., Tu, S. \& Shi, J. Time-weighted LSTM model with redefined labeling for stock trend prediction. {\em 2017 IEEE 29th International Conference On Tools With Artificial Intelligence (ICTAI)}. pp. 1210-1217 (2017)

\bibitem{zhang2018novel}
Zhang, J., Cui, S., Xu, Y., Li, Q. \& Li, T. A novel data-driven stock price trend prediction system. {\em Expert Systems With Applications}. 97 pp. 60-69 (2018)

\bibitem{zou2022astock}
J.~Zou, H.~Cao, L.~Liu, Y.~Lin, E.~Abbasnejad, and J.~Q. Shi, ``Astock: A new dataset and automated stock trading based on stock-specific news analyzing model,'' \emph{arXiv preprint arXiv:2206.06606}, 2022.

\bibitem{chen2021graph}
Chen, Q. \& Robert, C. Graph-based learning for stock movement prediction with textual and relational data. {\em ArXiv Preprint ArXiv:2107.10941}. (2021)

\bibitem{pawlowski2023effective}
Pawłowski, M., Wróblewska, A. \& Sysko-Romańczuk, S. Effective Techniques for Multimodal Data Fusion: A Comparative Analysis. {\em Sensors}. 23, 2381 (2023)

\bibitem{fukui2016multimodal}
Fukui, A., Park, D., Yang, D., Rohrbach, A., Darrell, T. \& Rohrbach, M. Multimodal compact bilinear pooling for visual question answering and visual grounding. {\em ArXiv Preprint ArXiv:1606.01847}. (2016)

\bibitem{wirojwatanakul2019multi}
Wirojwatanakul, P. \& Wangperawong, A. Multi-label product categorization using multi-modal fusion models. {\em ArXiv Preprint ArXiv:1907.00420}. (2019)

\bibitem{singh2019towards}
Singh, A., Natarajan, V., Shah, M., Jiang, Y., Chen, X., Batra, D., Parikh, D. \& Rohrbach, M. Towards vqa models that can read. {\em Proceedings Of The IEEE/CVF Conference On Computer Vision And Pattern Recognition}. pp. 8317-8326 (2019)

\bibitem{zhang2018adaptive}
Zhang, Q., Fu, J., Liu, X. \& Huang, X. Adaptive co-attention network for named entity recognition in tweets. {\em Proceedings Of The AAAI Conference On Artificial Intelligence}. 32 (2018)

\bibitem{yu2019deep}
Yu, Z., Yu, J., Cui, Y., Tao, D. \& Tian, Q. Deep modular co-attention networks for visual question answering. {\em Proceedings Of The IEEE/CVF Conference On Computer Vision And Pattern Recognition}. pp. 6281-6290 (2019)

\bibitem{huynh2023efficient}
T.~T. Huynh, M.~H. Nguyen, T.~T. Nguyen, P.~L. Nguyen, M.~Weidlich, Q.~V.~H. Nguyen, and K.~Aberer, ``Efficient integration of multi-order dynamics and internal dynamics in stock movement prediction,'' in \emph{Proceedings of the Sixteenth ACM International Conference on Web Search and Data Mining}, 2023,
  pp. 850--858.

\bibitem{daiya2021stock}
Daiya, D. \& Lin, C. Stock movement prediction and portfolio management via multimodal learning with transformer. {\em ICASSP 2021-2021 IEEE International Conference On Acoustics, Speech And Signal Processing (ICASSP)}. pp. 3305-3309 (2021)

\bibitem{wang2023essential}
Wang, J., Hu, Y., Jiang, T., Tan, J. \& Li, Q. Essential tensor learning for multimodal information-driven stock movement prediction. {\em Knowledge-Based Systems}. 262 pp. 110262 (2023)

\bibitem{he2021multi}
S.~He and S.~Gu, ``Multi-modal attention network for stock movements prediction,'' \emph{arXiv preprint arXiv:2112.13593}, 2021.

\bibitem{zhi2020factorized}
Zhi-Xuan, T., Soh, H. \& Ong, D. Factorized inference in deep markov models for incomplete multimodal time series. {\em Proceedings Of The AAAI Conference On Artificial Intelligence}. 34, 10334-10341 (2020)

\bibitem{chen2020hgmf}
Chen, J. \& Zhang, A. Hgmf: heterogeneous graph-based fusion for multimodal data with incompleteness. {\em Proceedings Of The 26th ACM SIGKDD International Conference On Knowledge Discovery \& Data Mining}. pp. 1295-1305 (2020)

\bibitem{ma2022multimodal}Ma, M., Ren, J., Zhao, L., Testuggine, D. \& Peng, X. Are multimodal transformers robust to missing modality?. {\em Proceedings Of The IEEE/CVF Conference On Computer Vision And Pattern Recognition}. pp. 18177-18186 (2022)

\bibitem{amrani2021noise}
Amrani, E., Ben-Ari, R., Rotman, D. \& Bronstein, A. Noise estimation using density estimation for self-supervised multimodal learning. {\em Proceedings Of The AAAI Conference On Artificial Intelligence}. 35, 6644-6652 (2021)

\bibitem{mai2023multimodal}
Mai, S., Sun, Y., Xiong, A., Zeng, Y. \& Hu, H. Multimodal Boosting: Addressing Noisy Modalities and Identifying Modality Contribution. {\em IEEE Transactions On Multimedia}. (2023)

\bibitem{yoo2021accurate}
J.~Yoo, Y.~Soun, Y.-c. Park, and U.~Kang, ``Accurate multivariate stock movement prediction via data-axis transformer with multi-level contexts,'' in \emph{Proceedings of the 27th ACM SIGKDD Conference on Knowledge Discovery \& Data Mining}, 2021, pp. 2037--2045.

\bibitem{feng2018enhancing}
Feng, F., Chen, H., He, X., Ding, J., Sun, M. \& Chua, T. Enhancing stock movement prediction with adversarial training. {\em ArXiv Preprint ArXiv:1810.09936}. (2018)

\bibitem{hsu2021fingat}
Hsu, Y., Tsai, Y. \& Li, C. FinGAT: Financial Graph Attention Networks for Recommending Top- K  K Profitable Stocks. {\em IEEE Transactions On Knowledge And Data Engineering}. 35, 469-481 (2021)

\bibitem{sagala2020stock}
Sagala, T., Saputri, M., Mahendra, R. \& Budi, I. Stock price movement prediction using technical analysis and sentiment analysis. {\em Proceedings Of The 2020 2nd Asia Pacific Information Technology Conference}. pp. 123-127 (2020)

\bibitem{poria2017review}
Poria, S., Cambria, E., Bajpai, R. \& Hussain, A. A review of affective computing: From unimodal analysis to multimodal fusion. {\em Information Fusion}. 37 pp. 98-125 (2017)

\bibitem{d2015review}
D'mello, S. \& Kory, J. A review and meta-analysis of multimodal affect detection systems. {\em ACM Computing Surveys (CSUR)}. 47, 1-36 (2015)

\bibitem{fukui-etal-2016-multimodal}
Fukui, A., Park, D., Yang, D., Rohrbach, A., Darrell, T. \& Rohrbach, M. Multimodal Compact Bilinear Pooling for Visual Question Answering and Visual Grounding. {\em Proceedings Of The 2016 Conference On Empirical Methods In Natural Language Processing}. pp. 457-468 (2016,11), https://aclanthology.org/D16-1044

\bibitem{lu2017knowing}
Lu, J., Xiong, C., Parikh, D. \& Socher, R. Knowing when to look: Adaptive attention via a visual sentinel for image captioning. {\em Proceedings Of The IEEE Conference On Computer Vision And Pattern Recognition}. pp. 375-383 (2017)

\bibitem{yu2018beyond}
Yu, Z., Yu, J., Xiang, C., Fan, J. \& Tao, D. Beyond bilinear: Generalized multimodal factorized high-order pooling for visual question answering. {\em IEEE Transactions On Neural Networks And Learning Systems}. 29, 5947-5959 (2018)

\bibitem{li2019beyond}
Li, X., Song, J., Gao, L., Liu, X., Huang, W., He, X. \& Gan, C. Beyond rnns: Positional self-attention with co-attention for video question answering. {\em Proceedings Of The AAAI Conference On Artificial Intelligence}. 33, 8658-8665 (2019)

\bibitem{zheng2023mmkgr}
Zheng, S., Wang, W., Qu, J., Yin, H., Chen, W. \& Zhao, L. Mmkgr: Multi-hop multi-modal knowledge graph reasoning. {\em 2023 IEEE 39th International Conference On Data Engineering (ICDE)}. pp. 96-109 (2023)

\bibitem{rajan2022cross}
Rajan, V., Brutti, A. \& Cavallaro, A. Is cross-attention preferable to self-attention for multi-modal emotion recognition?. {\em ICASSP 2022-2022 IEEE International Conference On Acoustics, Speech And Signal Processing (ICASSP)}. pp. 4693-4697 (2022)

\bibitem{li2021selfdoc}
Li, P., Gu, J., Kuen, J., Morariu, V., Zhao, H., Jain, R., Manjunatha, V. \& Liu, H. Selfdoc: Self-supervised document representation learning. {\em Proceedings Of The IEEE/CVF Conference On Computer Vision And Pattern Recognition}. pp. 5652-5660 (2021)

\bibitem{jaegle2021perceiver}
Jaegle, A., Borgeaud, S., Alayrac, J., Doersch, C., Ionescu, C., Ding, D., Koppula, S., Zoran, D., Brock, A., Shelhamer, E. \& Others Perceiver io: A general architecture for structured inputs \& outputs. {\em ArXiv Preprint ArXiv:2107.14795}. (2021)

\bibitem{zhou2023transformer}
Zhou, H., Yu, Y., Wang, C., Zhang, S., Gao, Y., Pan, J., Shao, J., Lu, G., Zhang, K. \& Li, W. A transformer-based representation-learning model with unified processing of multimodal input for clinical diagnostics. {\em Nature Biomedical Engineering}. pp. 1-13 (2023)

\bibitem{tsai2022multimodal}
Tsai, J. \& Chu, W. Multimodal Fusion with Cross-Modal Attention for Action Recognition in Still Images. {\em Proceedings Of The 4th ACM International Conference On Multimedia In Asia}. pp. 1-5 (2022)

\bibitem{jozefowicz2016exploring}
Jozefowicz, R., Vinyals, O., Schuster, M., Shazeer, N. \& Wu, Y. Exploring the limits of language modeling. {\em ArXiv Preprint ArXiv:1602.02410}. (2016)

\bibitem{dauphin2017language}
Dauphin, Y., Fan, A., Auli, M. \& Grangier, D. Language modeling with gated convolutional networks. {\em International Conference On Machine Learning}. pp. 933-941 (2017)

\bibitem{hua2022transformer}
Hua, W., Dai, Z., Liu, H. \& Le, Q. Transformer quality in linear time. {\em International Conference On Machine Learning}. pp. 9099-9117 (2022)

\bibitem{ye2021multi}
Ye, J., Zhao, J., Ye, K. \& Xu, C. Multi-graph convolutional network for relationship-driven stock movement prediction. {\em 2020 25th International Conference On Pattern Recognition (ICPR)}. pp. 6702-6709 (2021)

\bibitem{touvron2023llama}
Touvron, H., Lavril, T., Izacard, G., Martinet, X., Lachaux, M., Lacroix, T., Rozière, B., Goyal, N., Hambro, E., Azhar, F. \& Others Llama: Open and efficient foundation language models. {\em ArXiv Preprint ArXiv:2302.13971}. (2023)

\bibitem{velickovic2017graph}
Velickovic, P., Cucurull, G., Casanova, A., Romero, A., Lio, P., Bengio, Y. \& Others Graph attention networks. {\em Stat}. 1050, 10-48550 (2017)

\bibitem{sak2014long}
Sak, H., Senior, A. \& Beaufays, F. Long short-term memory recurrent neural network architectures for large scale acoustic modeling.  (2014)

\bibitem{vaswani2017attention}
Vaswani, A., Shazeer, N., Parmar, N., Uszkoreit, J., Jones, L., Gomez, A., Kaiser, Ł. \& Polosukhin, I. Attention is all you need. {\em Advances In Neural Information Processing Systems}. 30 (2017)

\bibitem{wu2018hybrid}
Wu, H., Zhang, W., Shen, W. \& Wang, J. Hybrid deep sequential modeling for social text-driven stock prediction. {\em Proceedings Of The 27th ACM International Conference On Information And Knowledge Management}. pp. 1627-1630 (2018)

\bibitem{matthews1975comparison}
Matthews, B. Comparison of the predicted and observed secondary structure of T4 phage lysozyme. {\em Biochimica Et Biophysica Acta (BBA)-Protein Structure}. 405, 442-451 (1975)

\bibitem{qin2017dual}
Qin, Y., Song, D., Chen, H., Cheng, W., Jiang, G. \& Cottrell, G. A dual-stage attention-based recurrent neural network for time series prediction. {\em ArXiv Preprint ArXiv:1704.02971}. (2017)

\bibitem{mikolov2013efficient}
Mikolov, T., Chen, K., Corrado, G. \& Dean, J. Efficient estimation of word representations in vector space. {\em ArXiv Preprint ArXiv:1301.3781}. (2013)

\bibitem{devlin2018bert}
Devlin, J., Chang, M., Lee, K. \& Toutanova, K. Bert: Pre-training of deep bidirectional transformers for language understanding. {\em ArXiv Preprint ArXiv:1810.04805}. (2018)

\bibitem{perozzi2014deepwalk}
Perozzi, B., Al-Rfou, R. \& Skiena, S. Deepwalk: Online learning of social representations. {\em Proceedings Of The 20th ACM SIGKDD International Conference On Knowledge Discovery And Data Mining}. pp. 701-710 (2014)

\bibitem{kipf2016semi}
Kipf, T. \& Welling, M. Semi-supervised classification with graph convolutional networks. {\em ArXiv Preprint ArXiv:1609.02907}. (2016)

\bibitem{velivckovic2017graph}
Veličković, P., Cucurull, G., Casanova, A., Romero, A., Lio, P. \& Bengio, Y. Graph attention networks. {\em ArXiv Preprint ArXiv:1710.10903}. (2017)

\end{thebibliography}
\end{document}